\documentclass[manuscript]{aastex}	
\usepackage{epstopdf}			
\usepackage{graphicx,color}		
\usepackage{amssymb}			
\usepackage{color}			
\usepackage{url}			
\usepackage{amsmath}			
\usepackage{rotating}			
\usepackage{float}			
\usepackage{textcomp}			
\usepackage{times}
\usepackage[colorlinks=true,citecolor=blue]{hyperref}
\usepackage[english]{babel}
\usepackage{subfigure}

\shorttitle{Dynamics of on-Disk Plume }
\shortauthors{V. Pant et al.}

\begin{document}

\title{Dynamics of on-disk plumes as observed with Interface Region Imaging Spectrograph, Atmospheric Imaging Assembly and Helioseismic and Magnetic Imager}

\author{Vaibhav Pant$^{1}$,
Laurent Dolla$^{2}$,
Rakesh Mazumder$^{1,3}$
Dipankar Banerjee$^{1,3}$,
S. Krishna Prasad$^{4}$
Vemareddy Panditi$^{1}$}
 
\affil{$^{1}$Indian Institute of Astrophysics, Koramangala, Bangalore 560034, India. e-mail: \url{vaibhav@iiap.res.in}\\
$^{2}$Solar-Terrestrial Center of Excellence, Royal Observatory of Belgium, Avenue Circulaire 3, B-1180 Brussels, Belgium\\
$^{3}$Center of Excellence in Space Sciences India, IISER Kolkata, Mohanpur 741246, West Bengal, India\\
$^{4}$Astrophysics Research Centre, School of Mathematics and Physics, Queen's University Belfast, Belfast BT7 1NN, UK}

\begin{abstract}
We examine the role of small-scale transients in the formation and evolution of solar coronal
plumes. We study the dynamics of plume footpoints seen in the vicinity of a coronal hole using the
Atmospheric Imaging Assembly (AIA) images,
the Helioseismic and Magnetic Imager (HMI) magnetogram on board the Solar Dynamics Observatory (SDO) 
and spectroscopic data from the Interface Region Imaging Spectrograph (IRIS). 
Quasi-periodic brightenings are  observed in the base of the plumes and are associated with magnetic flux changes. 
With the high spectral and spatial resolution of IRIS, we identify the sources of these oscillations and try to understand what role the transients at the foot points can play
in sustaining the coronal plumes.
IRIS sit and stare observation provide a unique opportunity to study the evolution of foot points of the plumes. 
We notice enhanced line width, intensity and large deviation from the average Doppler shift in the line profiles at specific instances which indicate the presence of flows at the foot points of plumes.
We propose that outflows (jet-like features) as a result of small scale reconnections affect the line profiles. These jet-like features may be also responsible for the generation of propagating disturbances within the plumes which are observed to be propagating to larger distances as recorded from multiple AIA channels. These propagating disturbances can be explained in terms of slow magnetoacoustic waves. 
\end{abstract}



\section{Introduction}
Coronal plumes, extending as bright narrow structures from the solar chromosphere into the high corona are mostly rooted in Coronal Holes (CH) or their neighbourhood \citep{1950BAN....11..150V,1958PASJ...10...49S,1965PASJ...17....1S,2001ApJ...560..490D}. They are plasma density enhancements in the low and extended corona aligned along the magnetic field \citep{2011A&ARv..19...35W,2011ApJ...736...32V}. Plume foot points are typically 4\arcsec~ wide and plumes expand rapidly with height \citep{1997SoPh..175..393D}. \citet{2006ESASP.617E..16R,2007ApJ...658..643R} showed that plumes are more than 5 times denser than the inter-plume regions at the base of the corona. Multi-wavelength UV-EUV imaging and spectral observations from Solar and Heliospheric Observatory (SoHO), Solar TErrestrial RElations Observatory (STEREO) and Solar Dynamic Observatory (SDO) allowed us to study the formation and evolution of plumes \citep[see review by][]{2011A&ARv..19...35W}. Plumes are difficult to observe in low latitude CHs because of the bright foreground and background emission \citep{2008SoPh..249...17W}. The low-latitude plumes appear to be
similar to their polar counterparts \citep{2011ApJ...736..130T}. \citet{1957PASJ....9..106S} first pointed out the association of polar plumes with magnetic flux concentrations.
It was also revealed by many studies that polar plumes arise from unipolar magnetic regions associated with the
supergranular network boundaries \citep{1965ApJ...141..832H,1968SoPh....3..321N,1995ApJ...447L.139F,1997SoPh..175..393D,2001ApJ...560..490D,1999A&A...350..286Y}
Coronal plumes can be formed due to magnetic reconnection of newly emerging magnetic flux
with the pre-existing dominant unipolar fields \citep{2011ApJ...727...30G}, which can further lead to localised heating \citep{1995ApJ...452..457W}. \citet{2008ApJ...682L.137R} and \citet {2014ApJ...787..118R} discovered that coronal jets are the precursors of plumes. 

\citet{1997ApJ...491L.111O} reported quasi-periodic variations in polar coronal holes and conjectured compressive waves to be responsible for them. Quasi-periodic brightness variations in plumes have been observed with Extreme ultraviolet Imaging Telescope (EIT) by \citet{1998ApJ...501L.217D}. They found presence of Propagating Disturbances (PDs) in several plumes with periods between 10 min and 15 min and speeds between 75 km s$^{-1}$ and 150 km s$^{-1}$ in the height range from 0.01 R$_{s}$ to 0.2 R$_{s}$. They conjectured such PDs to be the compressive waves. A number of studies followed, reporting such oscillations in plumes, interplumes, and coronal holes, using spectroscopic data obtained with Coronal Diagnostic Spectrometer (CDS) and Solar Ultraviolet Measurements of Emitted Radiation (SUMER) onboard SOHO and Extreme
ultraviolet Imaging Spectrometer (EIS) onboard Hinode \citep{2001A&A...377..691B,2000SoPh..196...63B,2006A&A...452.1059O,2009A&A...499L..29B,2009A&A...493..251G}
. Using Extreme UltraViolet (EUV) observations from STEREO, \citet{2010A&A...510L...2M} analysed several plume structures and suggested
that PDs along polar plumes could be due to the collimated
high-speed plasma jets that have similar properties as slow magnetoacoustic waves. Further, they conjectured that these jets could be responsible for loading a significant amount of heated plasma into the fast solar wind. Indeed, recent high-resolution observations have revealed jet-like flows at the bases of plumes \citep{2014ApJ...787..118R} and at network boundaries in coronal holes \citep{2014Sci...346A.315T}. The plume formation and dynamics are still a matter of debate related to the broader issue of wave propagation, plasma jets and their role in the acceleration of the fast solar wind.

To understand the nature of the source regions of plumes and the PDs, we study the foot-points of an on-disk plume adjacent to a coronal hole as seen in the Atmospheric Imaging Assembly (AIA) EUV coronal images \citep{2012SoPh..275...17L} and in the Helioseismic and 
Magnetic Imager (HMI) magnetogram \citep{2012SoPh..275..229S} on the Solar Dynamics Observatory 
and simultaneously with Interface Region Imaging Spectrograph \citep[IRIS;][]{2014SoPh..289.2733D}.
Combining  imaging and spectroscopic observations, we focus on the dynamics of this plume. We show that the jet-like features (termed as jetlets) load mass to the plumes and can be also responsible for the generation of the PDs as observed at higher heights.

\section{Data analysis and Results}

\subsection{Observation and Data Reduction}
\begin{figure*}
\centering
\subfigure[]{\includegraphics[angle=0,scale=.35]{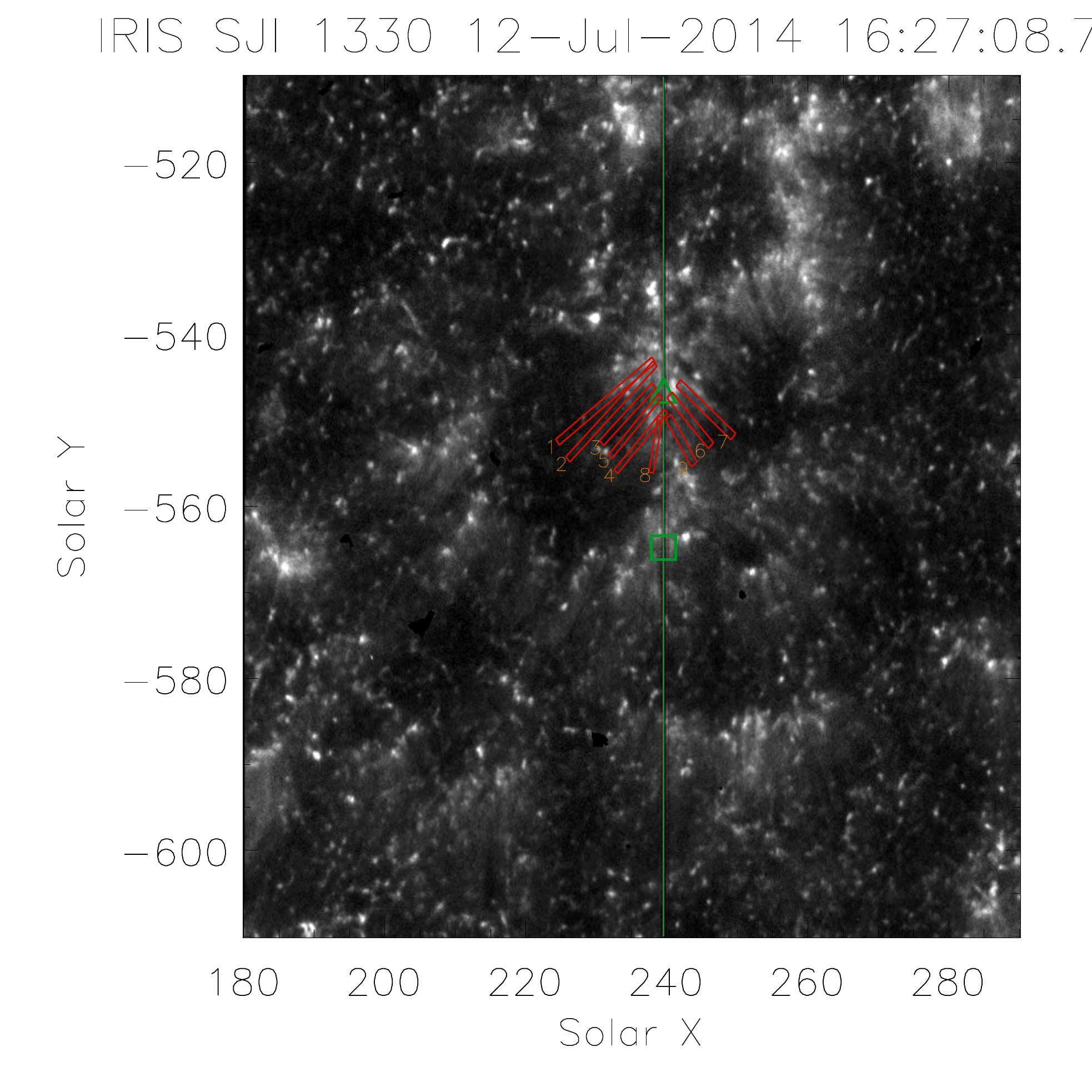}}
\subfigure[]{\includegraphics[angle=0,scale=.35]{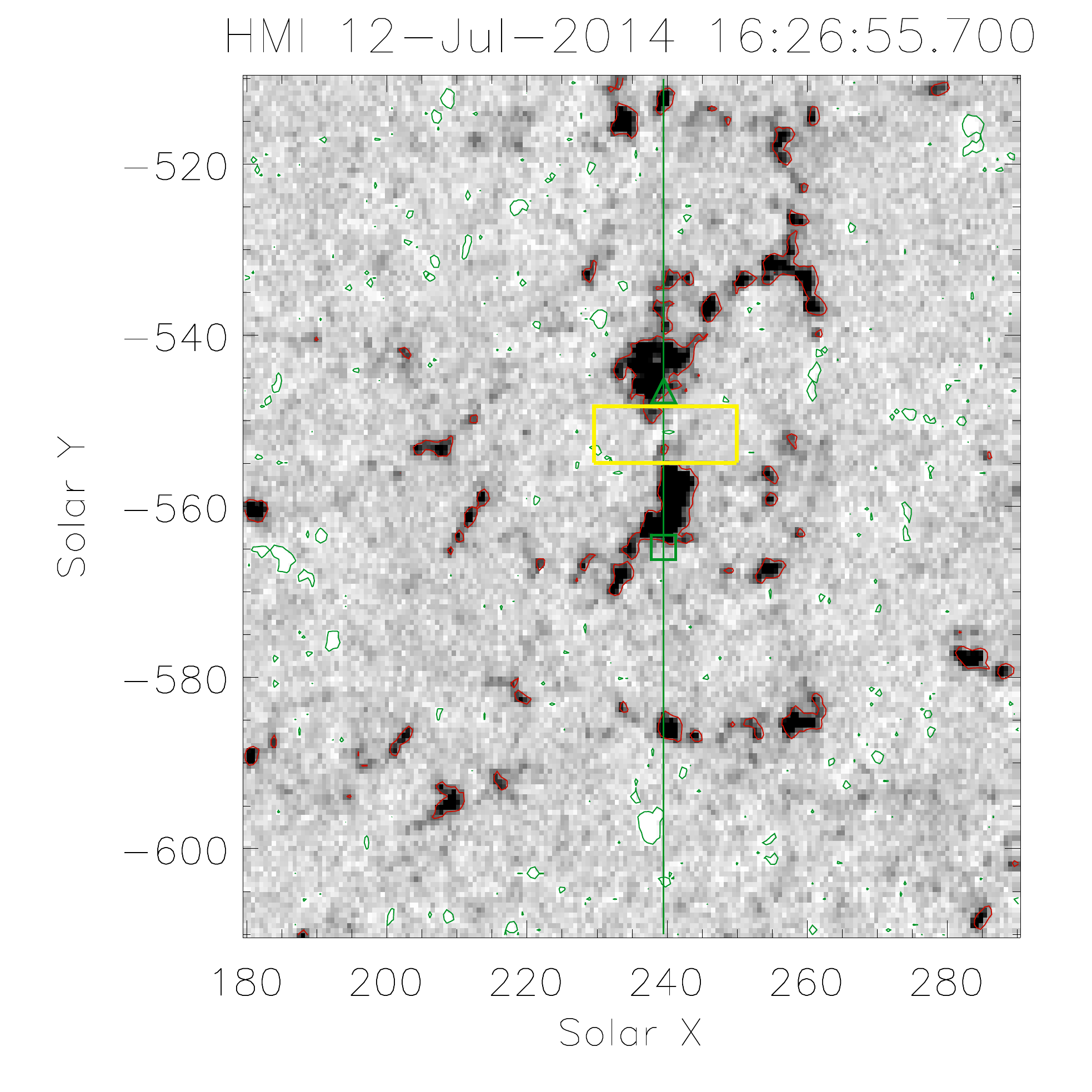}}
\subfigure[]{\includegraphics[angle=0,scale=.35]{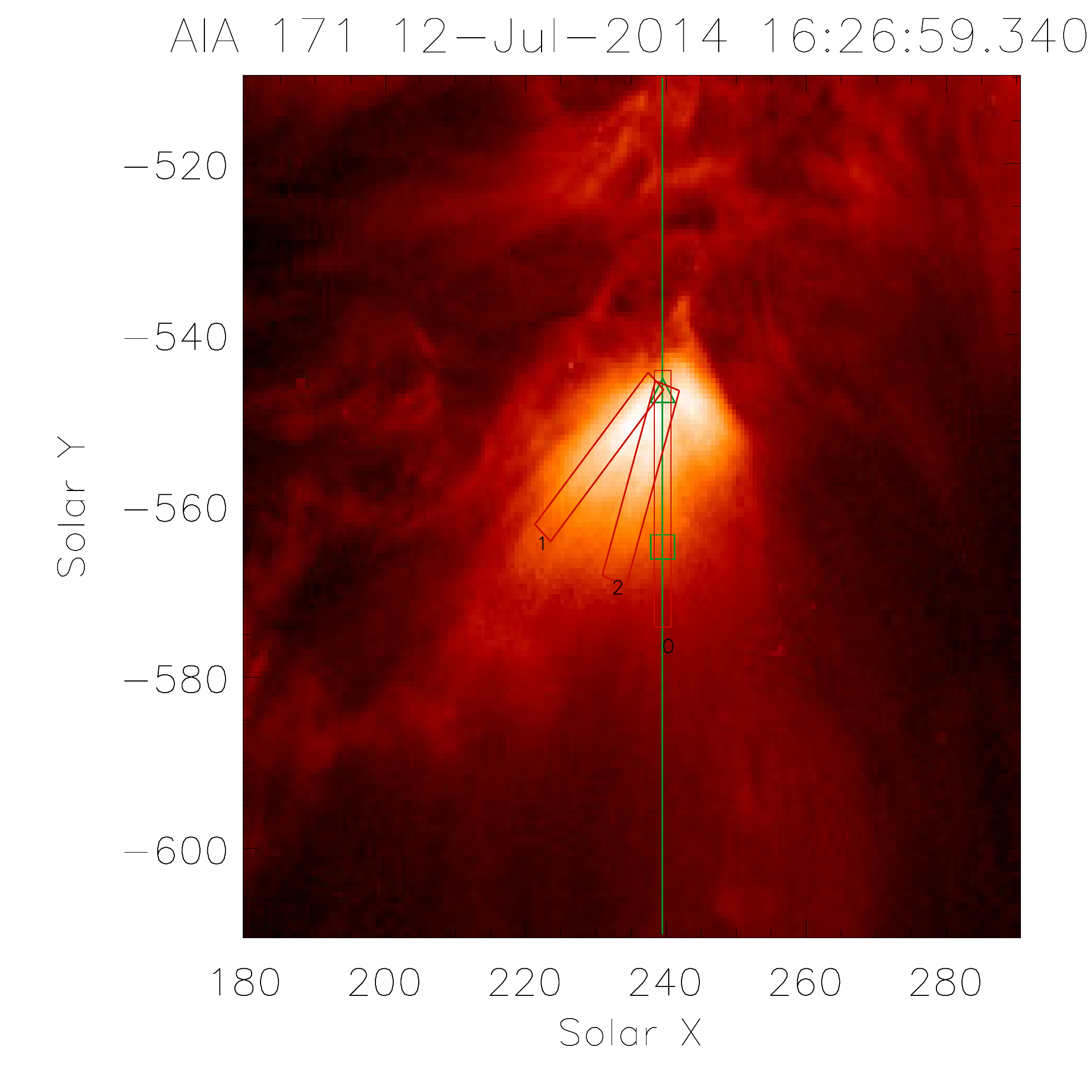}}
\subfigure[]{\includegraphics[angle=0,scale=.35]{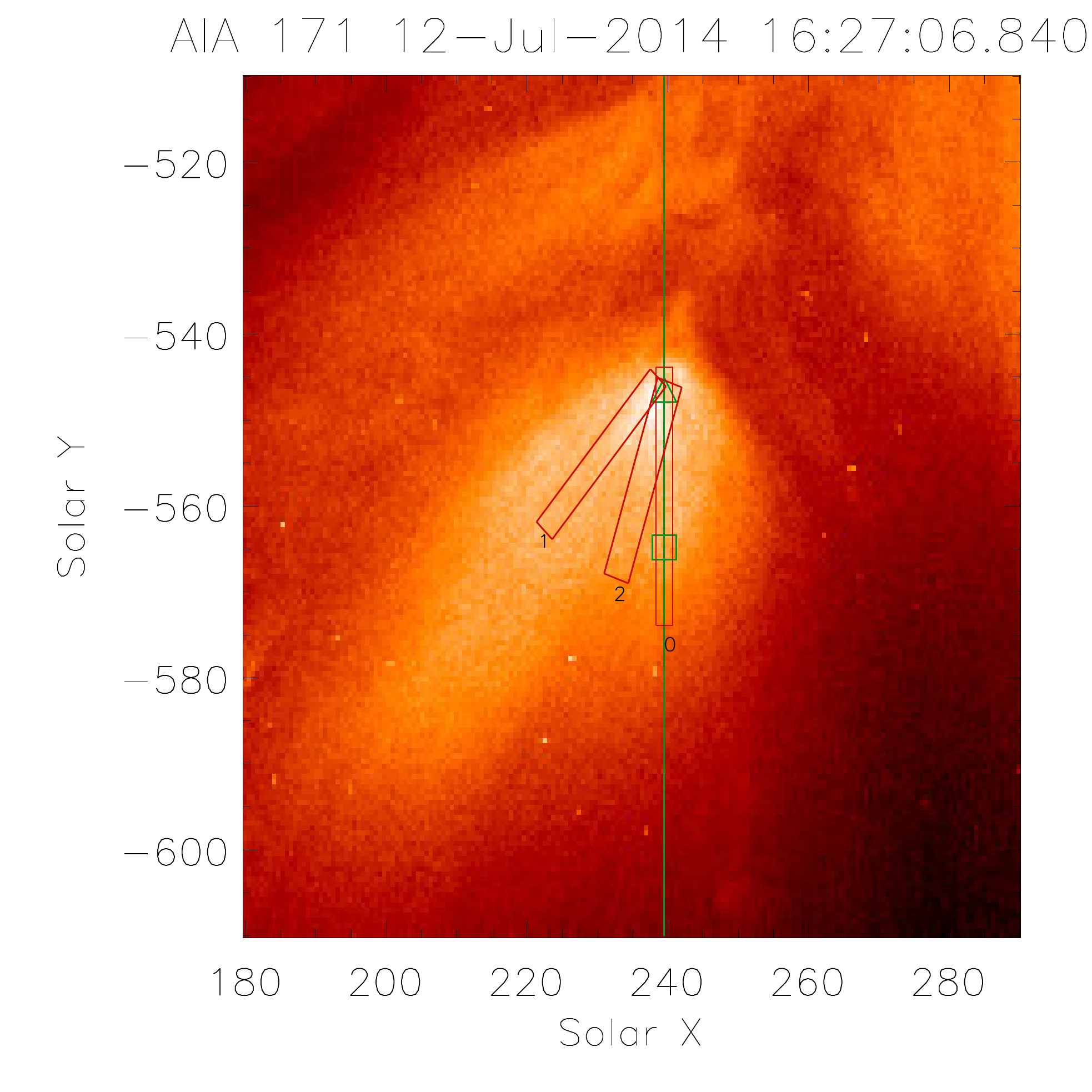}}
\caption{\textrm{(a)} : IRIS 1330 \r{A} SJI represents our Region Of Interest (ROI), 
covering an on-disk plume adjacent to a coronal hole. Nine artificial slices are placed for further analysis. \textrm{(b)}: HMI LOS magnetogram. The yellow box represents the region used for calculating the positive flux (see Fig.~\ref{fig4}). \textrm{(c)}: AIA/SDO 171 \r{A} image overplotted with three artificial slices used to create time-distance maps (see Fig.~\ref{fig5}). \textrm{(d)}: Same as (c) for AIA/SDO 193 \r{A} image. The vertical green line on each image represents the position of the IRIS slit. The green triangle is the location where we study the variation of different parameters of the \ion{Si}{4} 1402.77 \r{A} line (see Fig.~\ref{fig4}). The green triangle and square symbols represent the `Y' positions along the IRIS slit used to analyze spectra (see Fig.~\ref{fig3} and Movie~2).}
\label{fig1} 
\end{figure*}

Observational data was obtained from IRIS, AIA and HMI instruments from 16:26~UT to 17:51~UT on 12 July 2014. 
We used AIA passbands at 193~\r{A}, 171~\r{A}, 1600~\r{A} and HMI line of sight (LOS) magnetograms. 
We used standard AIA prep routines to produce level~1.5 images. IRIS data is taken in sit-and-stare mode with the slit centered at $239'', -559''$ pointing adjacent to a coronal hole. This dataset has three spectral lines namely, \ion{C}{2} 1335.71~\r{A}, \ion{Si}{4} 1402.77~\r{A} and \ion{Mg}{2} k 2796~\r{A}. We use \ion{Si}{4} 1402.77~\r{A} for this study.
Slit-jaw images (SJIs) were available only with the 1330~\r{A} filter. We have used  IRIS Level~2 processed data which is corrected for dark current, flat field and geometrical corrections etc.
The cadence of 1330~\r{A} SJIs and spectra was $\sim$5 sec with a pixel size of $0.166''$.
AIA and HMI images were de-rotated before AIA 1600~\r{A} images were co-aligned with  IRIS SJIs 1330~\r{A} using cross-correlation.
Fig.~\ref{fig1} shows the plume as seen in various AIA channels, IRIS 1330~\r{A} SJI and HMI LOS magnetogram.  
The magnetogram (Fig.~\ref{fig1} (b)) shows that the plume is dominated by negative polarity magnetic field marked with red contours (-50 G). We made an 84-min Movie~1 (available online) with HMI LOS magnetogram images, which shows the evolution of negative and positive flux with time. In the movie, red and green contours represent the magnetic field strength of -50 G and 20 G respectively.
  
\begin{figure*}[h]
\centering
\subfigure[]{\includegraphics[angle=0,scale=.4]{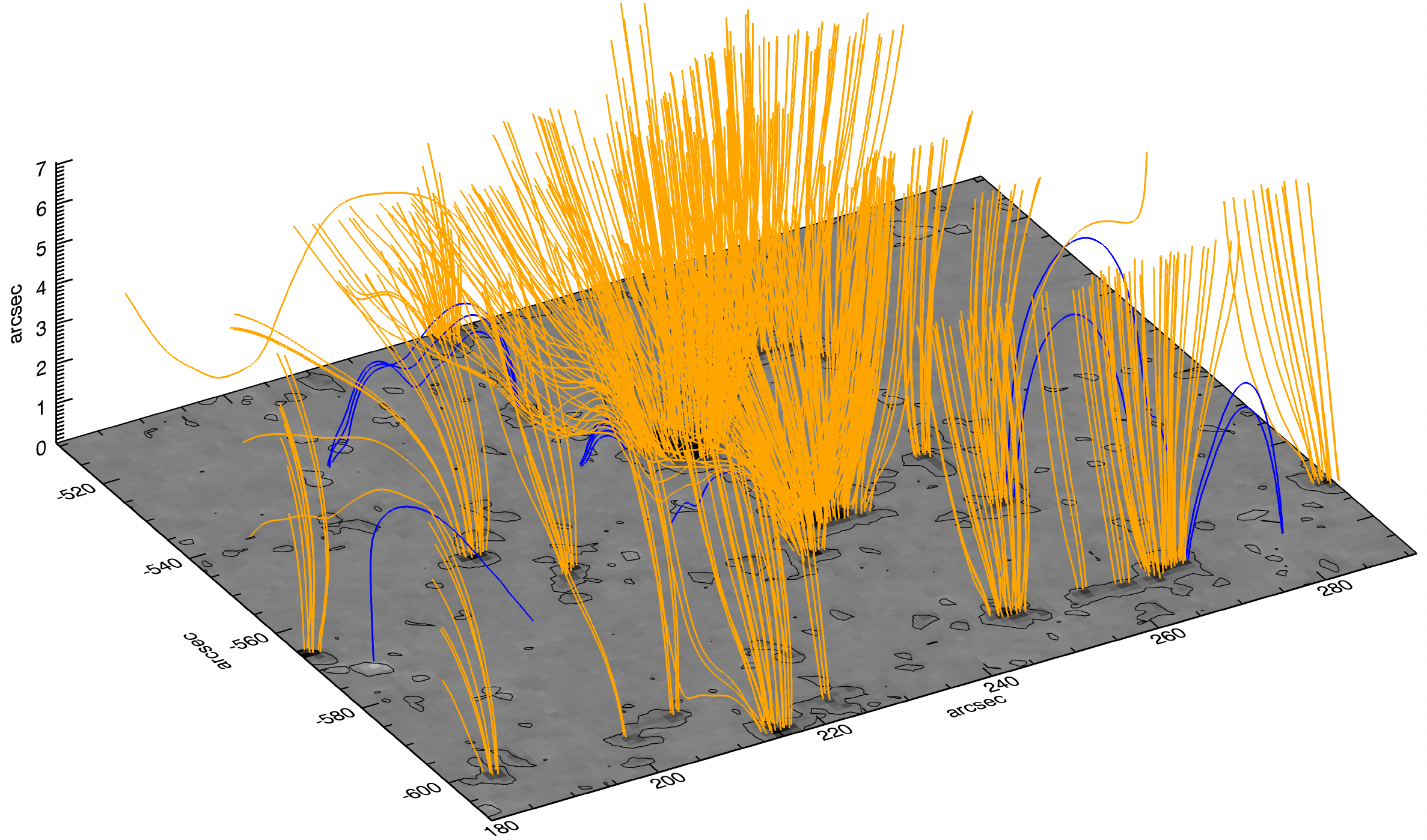}}
\subfigure[]{\includegraphics[angle=0,scale=0.4]{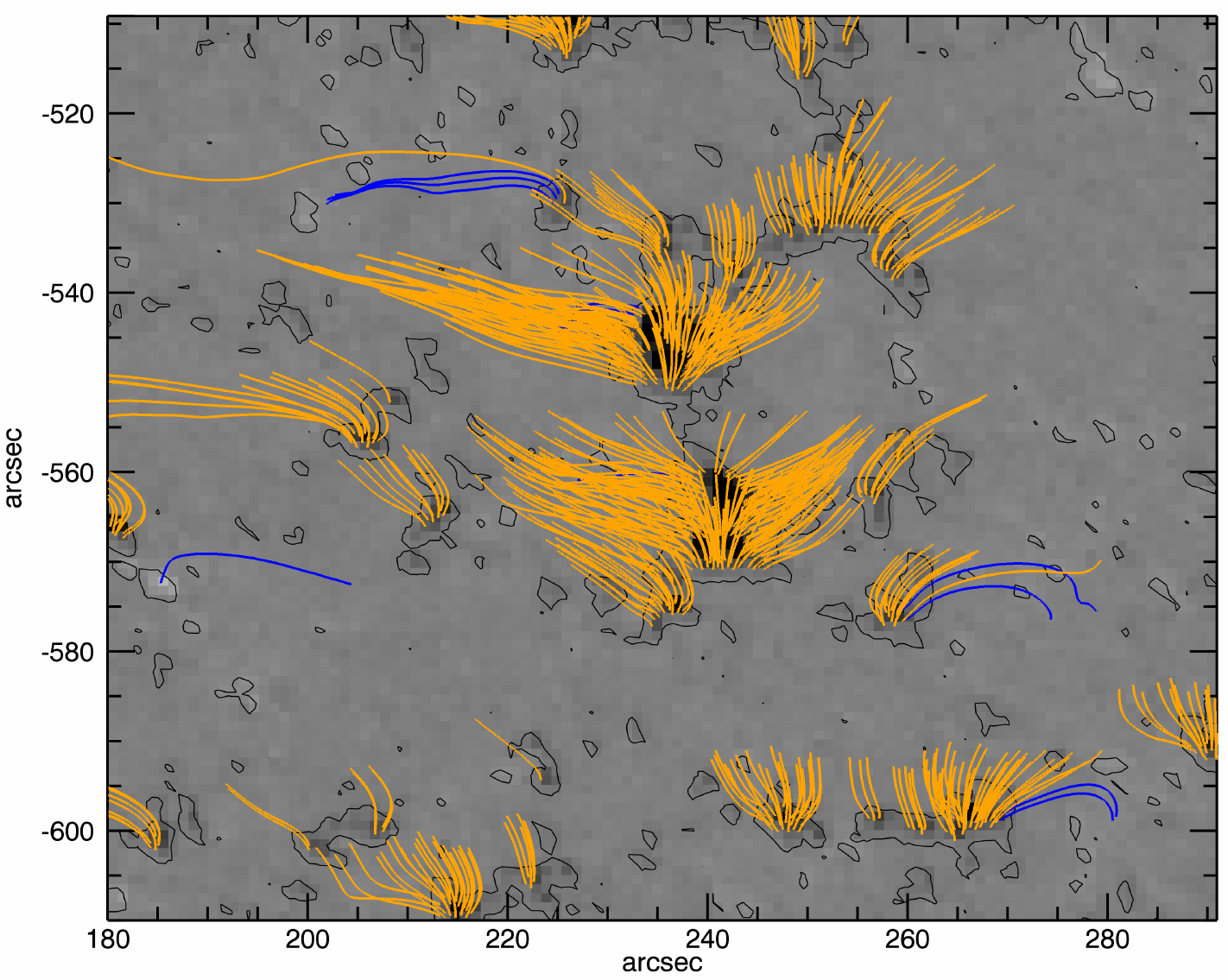}}
\caption{\textrm{(a)}: 3D view of extrapolated field lines. \textrm{(b)}: 2D view of extrapolated field lines.  }
\label{extrapol} 
\end{figure*}

Fig.~\ref{fig1} clearly shows that the IRIS slit is crossing the foot points of the plume.
We extrapolate the coronal magnetic field with a potential field approximation, using the LOS magnetic field observed by HMI as boundary condition at the photosphere (Fig.~\ref{extrapol}). We select a large field of view (FOV) around this plume structure to approximately satisfy flux balance condition. This FOV magnetogram is projected to disc centre correcting for the difference between normal and LOS values and we compute the potential field in a 3D box \citep{1989ApJS...69..323G} encompassing the plume structure.

We trace field lines having foot points in strong magnetic patches (${|B_{z}|} > $ 25 G). In Figs.~\ref{extrapol} (a) and (b), they are overlaid on the boundary magnetogram along with contours ($\pm$ 25 G). Most of the field lines from the magnetic regions are open (yellow). The shape of the two major open-field structure associated with the strong negative magnetic patches are consistent with the funnel structure of the plume originating from the network boundary, as revealed from the AIA images (Fig.~\ref{fig1} (c) and (d)). However, only the nothernmost structure appears bright in the AIA images at the time of the observations. Note that the difference in the orientation is due to projection which is corrected in this model. It is also noted that the plume originates from the network boundary, as revealed from IRIS SJIs (Fig.~\ref{fig1}(a)).\\

Apart from these open field lines, low lying field lines (blue) connects the weak field regions in the immediate neighbourhood of the main plume structure. Movie~1 shows the continuous emergence of positive flux followed by flux cancellation. Reconciling the model of \citet{1994ApJ...431L..51S}, the open-dipped field lines at the edges of the plume, in a dynamical scenario, facilitate continuous reconnection with the neighbouring closed field lines to supply heat and energy for sustaining the plume.

\subsection{Spectroscopic analysis}
\begin{figure*}
\centering
\includegraphics[scale=0.6]{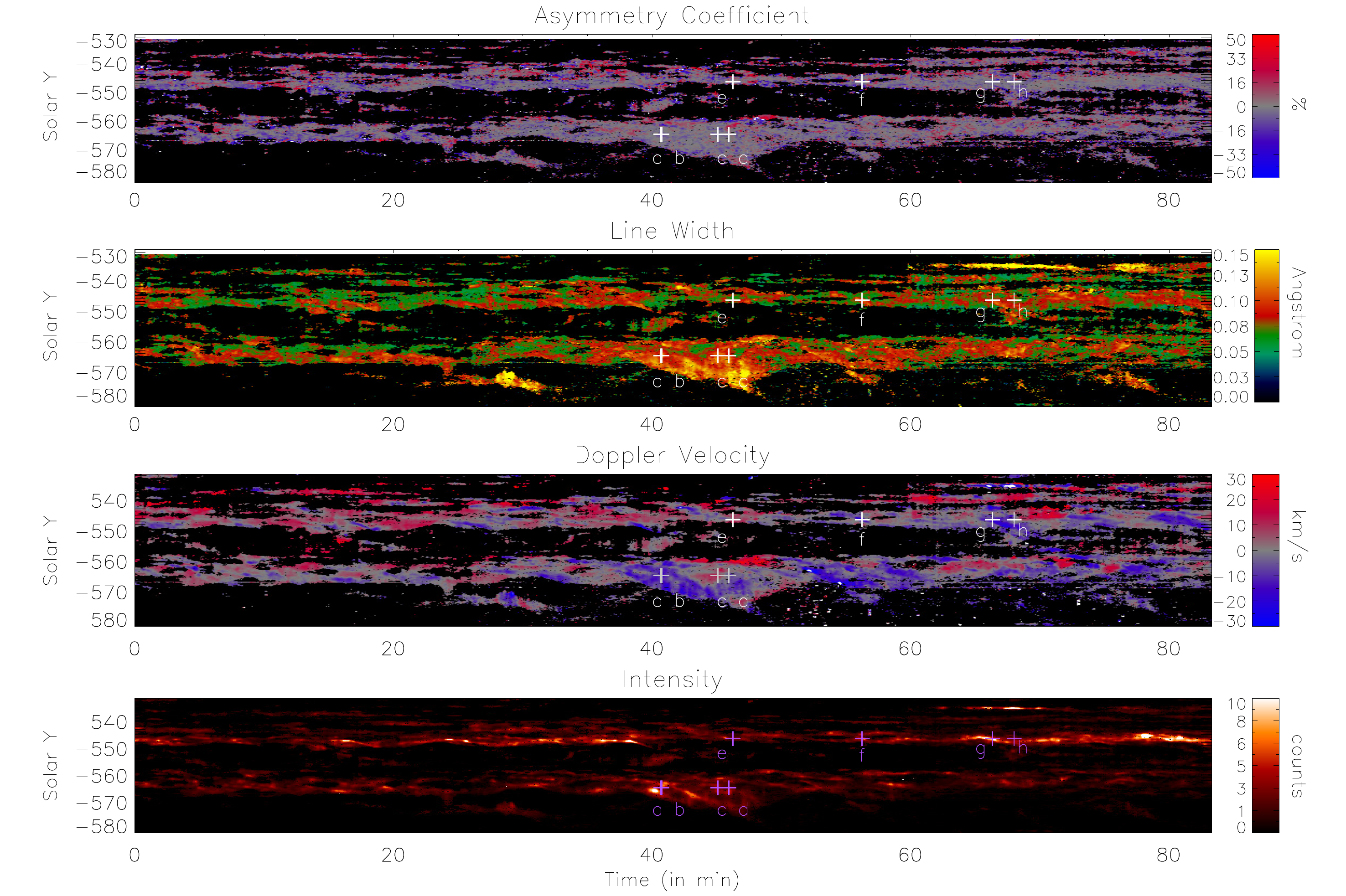}
\caption{Time evolution of peak intensity, Doppler velocity, line width and asymmetry coefficient as derived from single Gaussian fitting over \ion{Si}{4} 1402.77~\r{A} spectra. Positions marked as (a), (b), (c), (d), (e), (f), (g) and (h) are used to show individual spectral profiles. (see Fig.~\ref{fig3})}
\label{fig2} 
\end{figure*}

We have focused on the dynamics of two foot points (negative polarity) as marked by a triangle and square (at `Y' positions -546$''$ and -564$''$ respectively) in Fig.~\ref{fig1}. 
We study the time evolution of this small region as seen in IRIS \ion{Si}{4} 1402.77~\r{A} spectral profiles.
At first, a single Gaussian fit is performed on the weighted averaged (three pixels along the slit and three in time) spectral profile. We average the spectral profiles to increase the signal-to-noise ratio, using error bars as calculated by {\it iris\_pixel\_error.pro} as weights.
From the single Gaussian fit we derive peak intensity, Doppler shift, line width and asymmetry coefficient (see, Fig.~\ref{fig2}). We use the median of the centroid of the fitted Gaussians as the rest wavelength to estimate the Doppler velocity. The evolution of the spectral profile at these two positions as a function of time is shown in Movie~2 (available online). In this movie IRIS 1330~\r{A} SJIs are also included to show the jet-like features at network boundaries.
To analyze the non-Gaussian aspect of the spectral profiles, we use the coefficient of asymmetry defined in \cite{Dolla_and_Zhukov2011}. However, contrary to what was done in the latter reference, we do not normalize by the error bars of spectral intensities, because we are interested, here, in quantifying the asymmetry and not in assessing its statistical significance. The coefficient of asymmetry therefore provides the percentage of area in the profile that deviates from the fitted single Gaussian, according to the pattern defined by the following formula:

\begin{equation} \label{eq coeff asym}
A =  \frac{1}{I_0} \sum_{k} \epsilon(\lambda) \cdot \textrm{sgn}(\Delta_k(\lambda)) \cdot \Delta_k(\lambda),
\end{equation}

%
%
where $\lambda$ is the wavelength, $I_0$ the total intensity in the spectral line and $\Delta (\lambda) = s_k(\lambda)-f_k(\lambda)$ is the difference between the spectrum $s_k(\lambda)$ and the fitted Gaussian $f_k(\lambda)$, which are discretized on bin $k$. The contribution factor $\epsilon(\lambda)$ is defined as follows: 
\begin{equation} \label{eq contribution factor}
\epsilon(\lambda) = \begin{cases}
-1 & \textrm{ if } \lambda \in [\lambda_0 - 2 \, \sigma; \lambda_0 - \sigma) \\
1 & \textrm{ if } \lambda \in [\lambda_0 - \sigma; \lambda_0) \\
-1 & \textrm{ if } \lambda \in{} (\lambda_0; \lambda_0 + \sigma]\\
1 & \textrm{ if } \lambda \in{} (\lambda_0 + \sigma; \lambda_0 + 2 \, \sigma]\\
0 & \textrm{ otherwise, }
\end{cases}
\end{equation}
where $\lambda_0$ is the center and $\sigma$ the half-width at $1/ \sqrt{e}$ of the fitted Gaussian. 
The sign of the coefficient of asymmetry indicates in which wing of the profile most imbalance is present (negative and positive on the blue and red wings, respectively).  
Another method of quantifying the profile asymmetry is the Red-blue (RB) asymmetry method. RB technique has been widely used in both optically thick \citep[e.g., in H-$\alpha$, ][] {2009ApJ...701..253M,2014ApJ...797...88H} and optically thin \citep[e.g., ][]{2009ApJ...701L...1D,2011ApJ...738...18T} spectral-line analysis.

\begin{figure}[h]
\centering
\includegraphics[width=15cm]{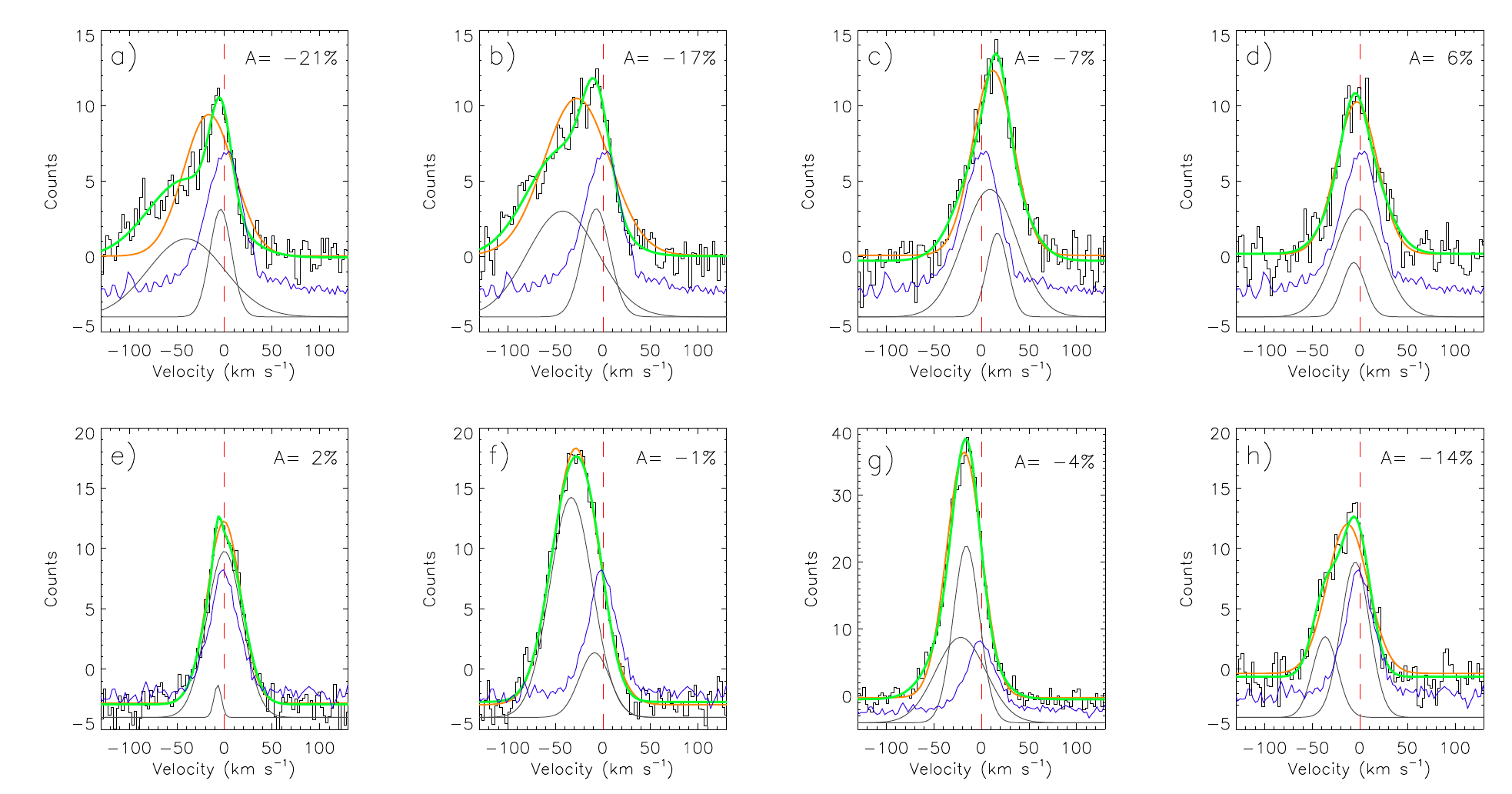}

\caption{Representative line profiles at different positions as marked in Fig.~\ref{fig2}. Orange  and green curves represent the best fit single Gaussian and double Gaussian over spectra. Two components of double Gaussian fit are shown in grey and shifted to the bottom for better visiblity. Blue curve represents the average spectrum over the first 20 min of the observation at respective positions.}
\label{fig3} 
\end{figure}

\begin{figure*}
\centering
\includegraphics[angle=0,clip,width=14cm]{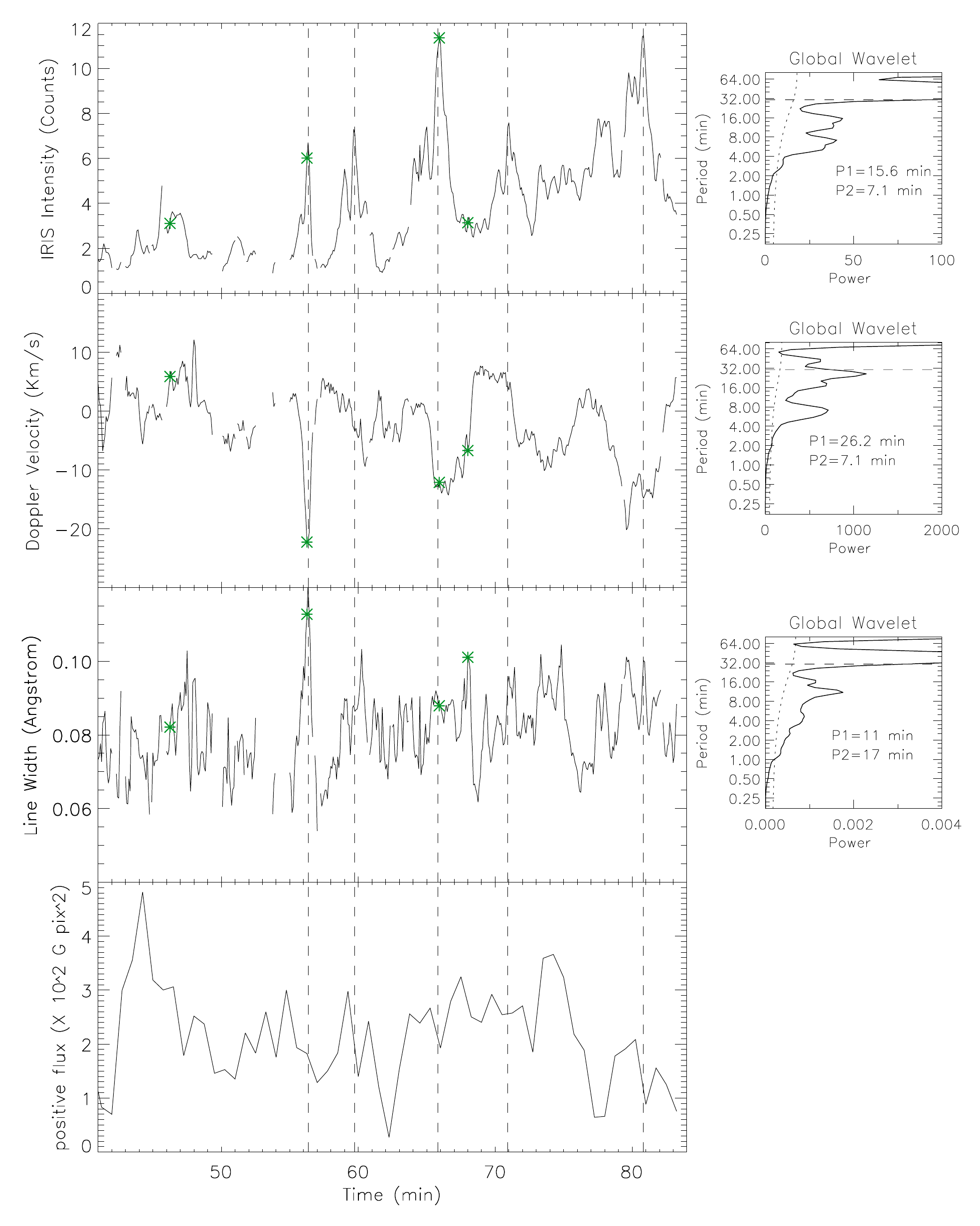}
\caption{{\it Left}: Time variation of the \ion{Si}{4} line parameters at the triangle location (IRIS) and of the positive magnetic flux within the yellow box (HMI), as shown in Fig.~\ref{fig1}. The green stars identify the times at which we show individual spectral profiles in Fig.~\ref{fig3}.
{\it Right}: Corresponding global wavelet power spectra. The 99 \% significance levels are overplotted with dotted lines. The first two significant periods are indicated.}
\label{fig4} 
\end{figure*}

Fig.~\ref{fig3} shows spectral profiles at different instances (as marked in Fig.~\ref{fig2}) along the slit. We find that the spectral profiles are significantly non-Gaussian with two or more components present. We fit the spectral profiles at these instances with a single Gaussian (shown in orange) and double Gaussian (in green). Spectra averaged during the first 20 min of observations at the same `Y' positions are shown in blue. Both components of double Gaussian are shown in grey. We observe both blueward and redward asymmetries at different instances (see Fig.~\ref{fig3}). Large asymmetry and double Gaussian behaviour of spectral profile indicates the presence of flows at these instances. Such behaviour can also be attributed to waves \citep{2010ApJ...724L.194V} but we can discard this interpretation in our observations because the associated fluctuations in intensity are too large to correspond to linear waves only (see Fig.~\ref{fig4}). Non-Gaussian behaviour is present on several occasions during the time of observation of IRIS at the positions marked as triangle and square in Fig.~\ref{fig1} (see Movie 2 online). IRIS 1330 \r{A} SJIs show the presence of several small-scale jets throughout the time of observation at these two positions. Reconnection jets cause enhancements at the line wings, which lead to large asymmetries, larger line width, larger intensity and large Doppler shift. The red shift in spectra could be a projection effect as the plume is close to the South pole. Besides this the redshifts may indeed imply that at least part of the material is falling back. This is consistent with the Fig.~\ref{fig4} where the brightenings are associated with strong blue shifts, with progressive evolution not only toward zero but also redshifts (i.e velocities slightly positive). There is probably a combination of both blue and red shifts during this recovery phase.\\

In Fig.~\ref{fig4} we show the wavelet analysis of spectral intensity, Doppler shift and line width at plume foot point marked in triangle (see, Fig.~\ref{fig1}). The missing values in the left panel of Fig.~\ref{fig4} correspond to the positions where a single Gaussian could not be fitted due to low counts. The green stars marks in Fig.~\ref{fig4} represent the instances we have chosen to show individual spectral profiles (see, Figs.~\ref{fig3} (e-h)). In the right panel global wavelet power peaks at 15.6 min and 7.1 min for intensity, 26.2 min and 7.1 min for Doppler velocity, and 11 min and 17 min for line width. A small peak $~\sim$16 min in global wavelet power of Doppler velocity is also present. The horizontal dashed line is the cutoff above which edge effects come into play, thus periods above the dashed line can not be trusted. Dotted line marks 99~\% significance level for a white noise process \citep{1998BAMS...79...61T}. \\
It is quite evident from Fig.~\ref{fig4} that the instances of the intensity peaks correspond to instances of large Doppler shifts and line widths and the flux cancellation of emerging positive flux. Movie 1 shows the evidence of positive polarity field around the dominant negative polarity. Cancellation of positive polarity at the edges of dominant negative polarity (region enclosed in yellow box in movie 1) leads to the formation of reconnection jets which are characterized by intensity enhancement in spectra and apparent outflows in IRIS 1330 \r{A} SJIs.\\
 
\subsection{Imaging Analysis}
\label{imagesection}
\begin{figure*}
\includegraphics[scale=.7,angle=90]{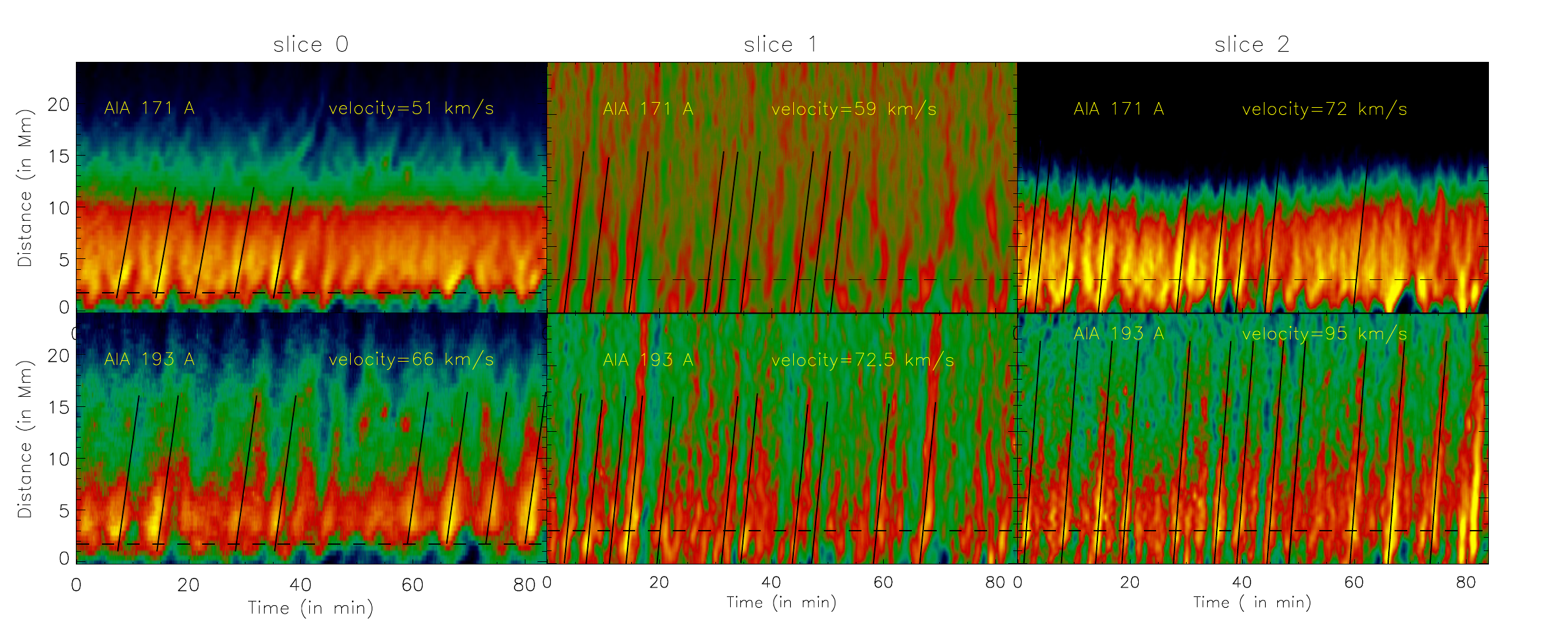}
\caption{ \textrm{(Top Panel)}: {\it Left}: Smooth background subtracted x-t map of slice 0 as shown in Fig.~\ref{fig1} (c) for AIA 171 \r{A}. Ridges are overplotted with best fitted straight line. {\it Middle}: Smooth background divided x-t map of slice 1 for AIA 171 \r{A}. Ridges are overplotted with best fitted straight line. {\it Right}: Same as (a) for slice 2. Dashed black line represents the position used for wavelet analysis as shown in Fig.~\ref{fig6}.
\textrm{(Bottom Panel)}: Same as Top panel for AIA 193~\r{A}.} 
\label{fig5} 
\end{figure*}

In order to compare dynamics at transition region (as seen from IRIS) and corona (as seen from AIA), we create smooth background subtracted time-distance maps of AIA 171 \r{A} and AIA 193 \r{A} for the artificial slice 0 (see Figs.~\ref{fig1} (c) and (d)) co-spatial with IRIS slit shown in green color in Fig.~\ref{fig1}. We choose the width of the slice to be four pixels. We average along the width of the slice to increase signal to noise. Resulting time-distance maps are shown in Fig.~\ref{fig5} (left panel). We observe quasi-periodic alternate bright and dark ridges. We fit the ridges with a straight line as done in \citet{2012SoPh..279..427K} and \citet{2012A&A...546A..50K}. We isolate a ridge and estimate the position of maximum intensity. We, then, fit the points of maximum intensity with a straight line, thus mean value of slope and error bars are estimated. The slope of the straight line gives an estimate of the speed of the PDs. Since we are interested in the average behaviour of the PD, we take the mean of the speeds of all the ridges and therefore some of the ridges in AIA 171~\r{A} seems to be more inclined than overplotted straight lines. We find the average velocity of PDs is 51$\pm$3 km~s$^{-1}$ and 66$\pm$8 km~s$^{-1}$ in AIA 171 \r{A} and 193 \r{A} respectively.\\
We carry out a similar analysis for two artificial slices marked as 1 and 2 in Figs.~\ref{fig1} (c) and (d) for AIA 171~\r{A} and 193~\r{A} respectively. The slices are placed by looking at the direction of propagation of significant PDs in unsharp mask images. Widths of slice 1 and 2 are chosen to be 5 and 6 pixels respectively. Corresponding time-distance maps with best fit overplotted ridges for slice 1 and 2 are shown in Fig.~\ref{fig5} (middle and right panel respectively).
We note the velocity is higher in the hotter channel (AIA 193 \r{A}) and the ratio between the velocity observed in AIA 193~\r{A} and 171~\r{A} is 1.29$\pm$0.23, 1.23$\pm$0.1 and 1.32$\pm$0.07 (as compared to the theoretical value of 1.25 if the PDs are magnetoacoustic waves) for slice 0, 1 and 2 respectively.\\
We also  estimate the velocities of PDs in slice 0, 1 and 2 using a cross-correlation method as done in \citet{2012SoPh..279..427K}. We isolate individual ridges and estimate the time lag using cross-correlation between two positions. Velocity of the ridge is estimated by dividing the distance between two positions by the time lag. Thus, we compute the velocities of several ridges and their mean value and standard deviation is estimated. Mean velocities and error bars (standard deviation for cross-correlation) using two different method for slice 0, 1 and 2 are summarized in Table.~\ref{tab1}.

\begin{deluxetable}{cccc}

\tablecolumns{4} 
\tablewidth{0pc} 
\tablecaption{Velocity of PDs using two different methods of ridge fitting} 
\tablehead{ 
\colhead{Slice}  & \colhead{Channel} &  \multicolumn{2}{c}{Velocity (km$s^{-1}$)}\\
\colhead{ } & \colhead{ } & \colhead{Using best fit straight line} & \colhead{Using correlation} \\
\cline{1-4}}
\startdata 
0 & AIA 171~\r{A} & 51$\pm$3 & 61$\pm$12\\
  & AIA 193~\r{A} & 66$\pm$8 & 81$\pm$11\\
1 & AIA 171~\r{A} & 59$\pm$3 & 60$\pm$14\\
  & AIA 193~\r{A} & 72.5$\pm$1 & 80$\pm$18\\
2 & AIA 171~\r{A} & 72$\pm$5 & 63$\pm$10\\
  & AIA 193~\r{A} & 95$\pm$3 & 83$\pm$25\\

\enddata 
\label{tab1}
\end{deluxetable}


\begin{deluxetable}{cccccc}

\tablecolumns{6} 
\tablewidth{0pc} 
\tablecaption{Dominiant Periodicity using wavelet transform} 
\tablehead{ 
\colhead{Slice}  & \colhead{Distance along slice} &  \multicolumn{2}{c}{AIA 171~\r{A}} & \multicolumn{2}{c}{AIA 193~\r{A}}\\
\colhead{ } & \colhead{ (Mm)} & \colhead{P1 (min)} & \colhead{P2 (min)} & \colhead{P1 (min)} & \colhead{P2 (min)}\\
\cline{1-6}}
\startdata 
0 & 2 & 14.4 & 7.2 & 7.9 & 14.4\\
  & 5 & 5.6 & 9.4 & 7.2 & 12.1 \\
  & 10 & 22.2 & 6.6 & 13.2 & 7.2\\
  & 15 & 12.1 & 7.2 & 12.1 & 7.2 \\
1 & 2.5 & 13.2 & 7.2 & 24.3 & 7.2\\
2 & 2.5 & 6.6 & -- & 26.4 & 6.6\\

\enddata 
\label{tab2}
\end{deluxetable}

\begin{figure*}
\includegraphics[scale=.7,angle=90]{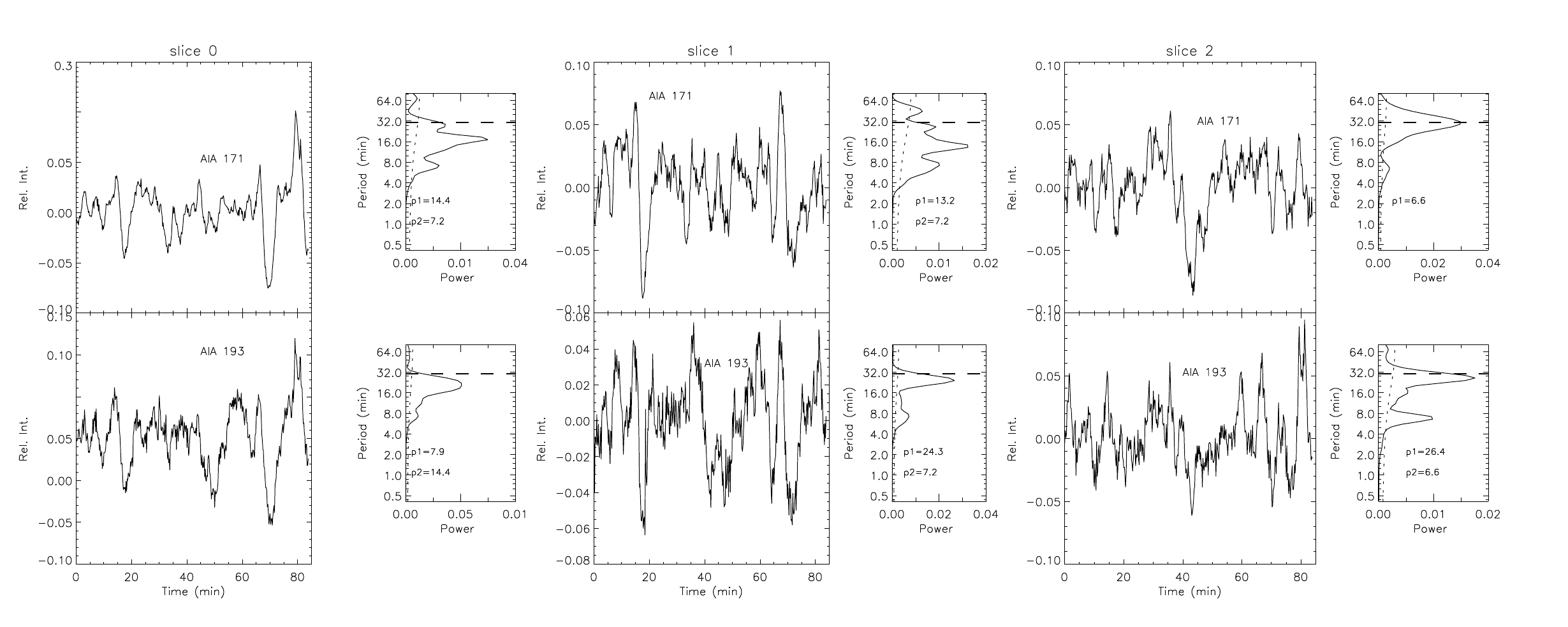}
\caption{ \textrm{(Top Panel)}: {\it Left}:  Detrended light curves corresponding to black dashed lines for slice 0 as shown in fig.~\ref{fig5} left panel. Corresponding global wavelet plots are shown with first two dominant periods. {\it Middle}: Same as Left panel for slice 1. {\it Right}: Same as Left panel for slice 2. \textrm{(Bottom Panel)}: Same as Top panel for AIA 193~\r{A}.}
\label{fig6} 
\end{figure*}

Since the velocity is higher in the hotter channel (AIA 193 \r{A}) using two different methods, we believe that these are propagating slow magnetoacoustic waves. We carry out a wavelet analysis at positions marked with dashed lines in Fig.~\ref{fig5} (left panel). We find that there exists at least two dominant significant periodicities close to the foot point of the plume (see  global wavelet power plot in Fig.~\ref{fig6}). We choose the significance level to be 99 \% for a white noise process. We find that the periodicities in intensity in AIA 171~\r{A} and 193 \r{A} at slice 0 are similar to those in IRIS \ion{Si}{4} 1402.77~\r{A} peak intensity (see, Fig.~\ref{fig4} top panel). We also find that the 7.2 min periodicity is present in both AIA 171 \r{A}, AIA 193 \r{A} and IRIS spectral intensity, which suggests that the quasi-periodic reconnection jet outflows could be responsible for quasi-periodic PDs seen in AIA 171 \r{A} and 193 \r{A}.\\

For slices 1 and 2 the rows marked with dashed lines in Fig.~\ref{fig5} are used in the wavelet analysis (Fig.~\ref{fig6}). We find the two dominant periodicities of 13.2 min and 7.2 min (24.3 min and 7.2 min) and 17.1 min and 6.6 min (26.4 min and 6.6 min) in AIA 171~\r{A} (193~\r{A}) for slices 1 and 2 respectively (see Table~\ref{tab2}).\\
We also carry out the wavelet analysis at other positions (10 Mm, 15 Mm in x-t map). The results are summarised in Table~\ref{tab2}.\\

\begin{figure*}

\subfigure[]{\includegraphics[bb=9 0 185 625,clip,angle=90,scale=.43]{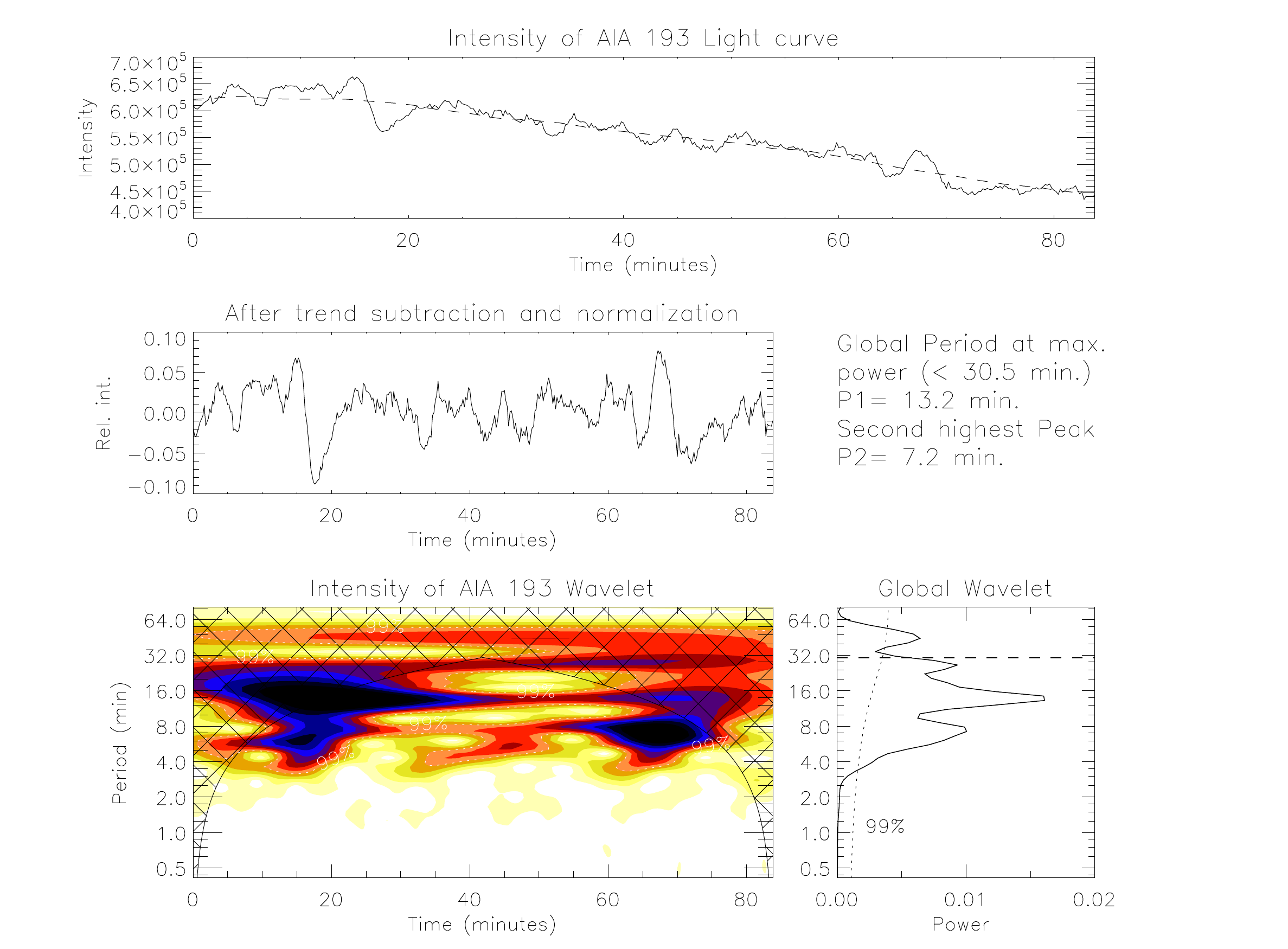}}
\subfigure[]{\includegraphics[bb=9 0 185 625,clip,angle=90,scale=.43]{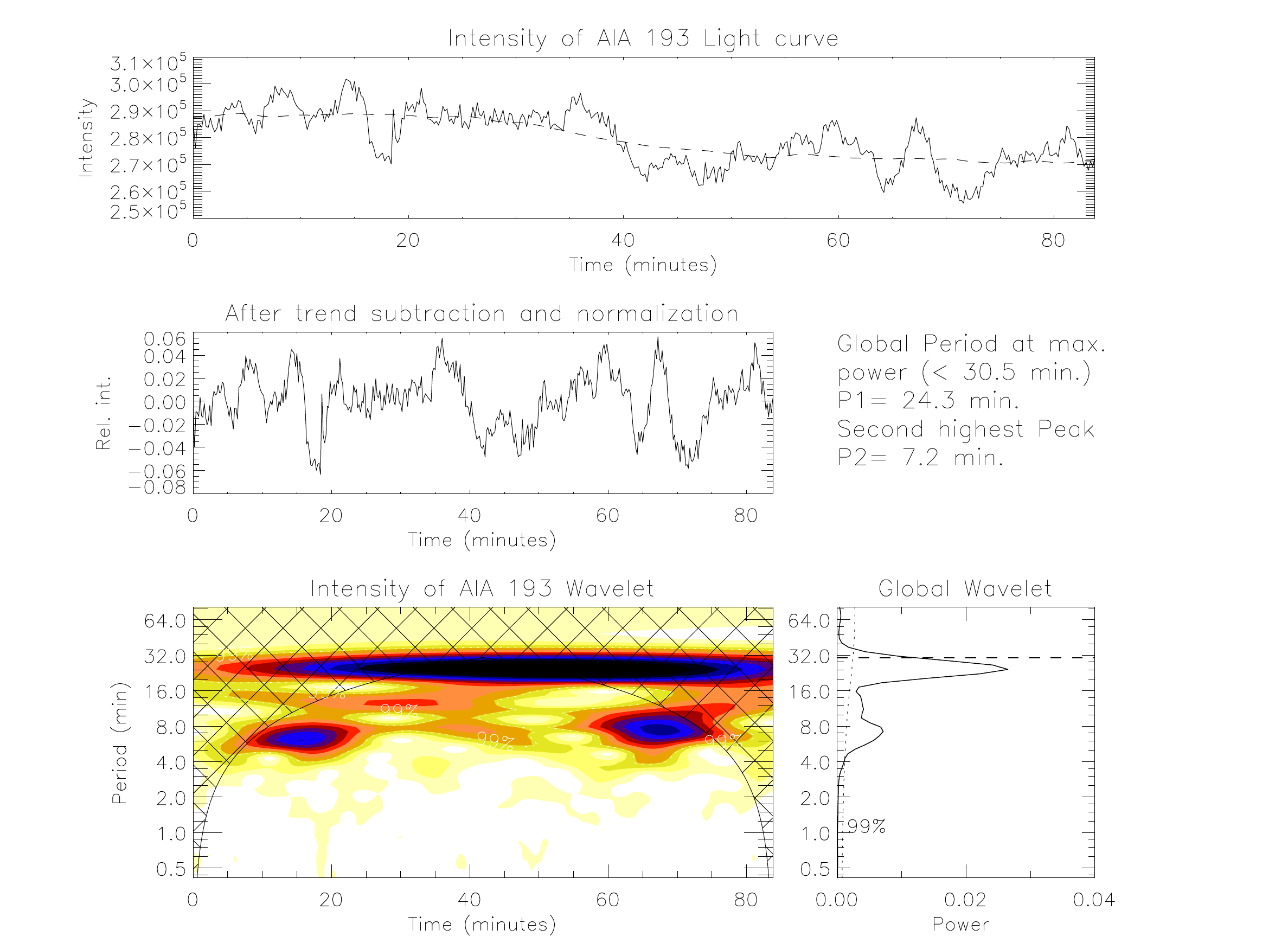}}
\subfigure[]{\includegraphics[bb=9 0 185 625,clip,angle=90,scale=.43]{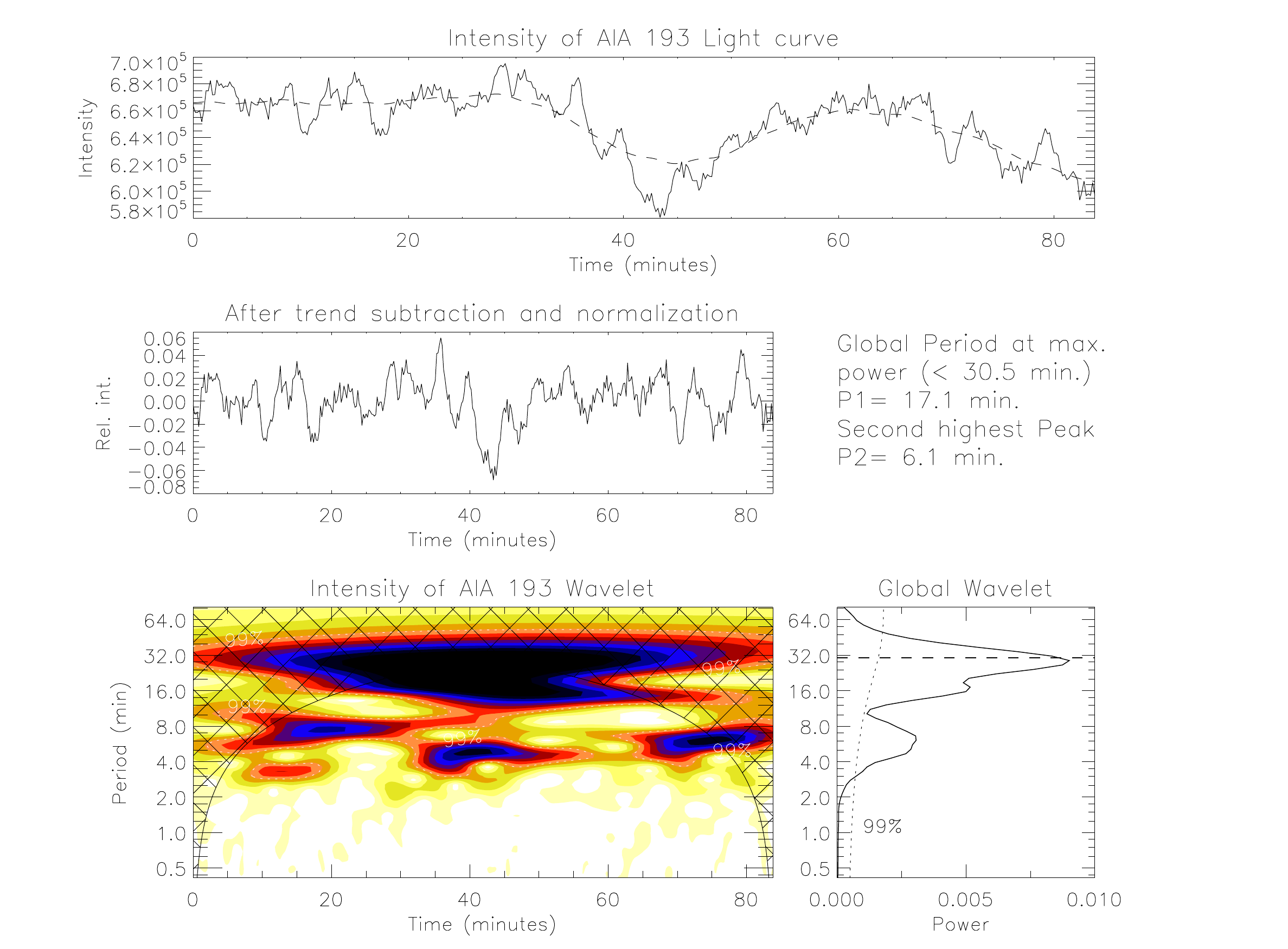}}
\subfigure[]{\includegraphics[bb=9 0 185 625,clip,angle=90,scale=.43]{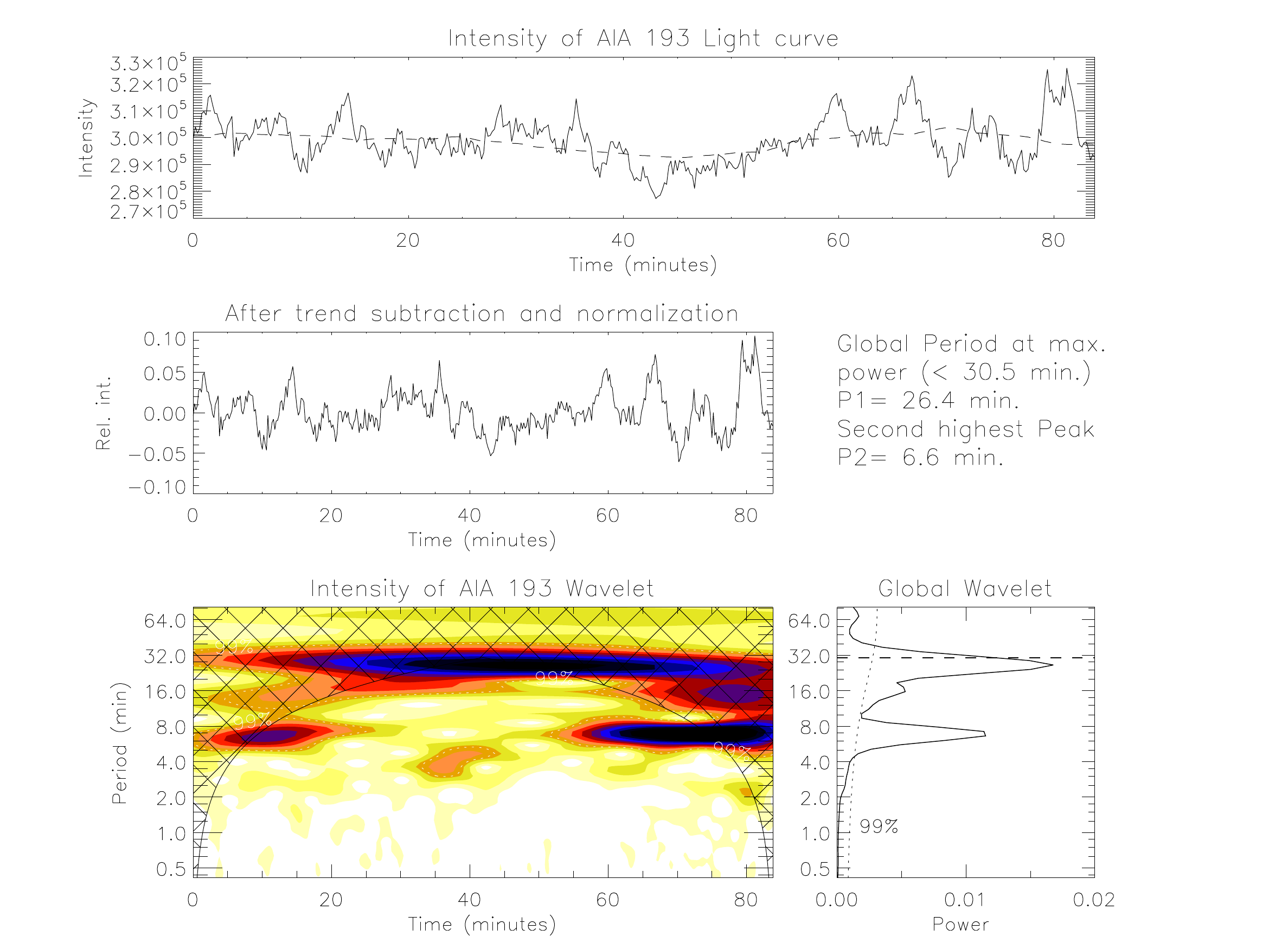}}
\caption{{\it (a)}: Wavelet map and global wavelet of AIA 171~\r{A} for slice 1 at position marked as dashed line in Fig.~\ref{fig5}. {\it (b)}: Same as {\it (a)} for AIA 193~\r{A}. {\it (c)}: Wavelet map of AIA 171~\r{A} for slice 2 at position marked as dashed line in Fig.~\ref{fig5}. {\it (d)}: Same as {\it (c)} for AIA 193~\r{A}.}
\label{wavelet} 
\end{figure*}

\begin{figure*}
\includegraphics[angle=90,scale=1]{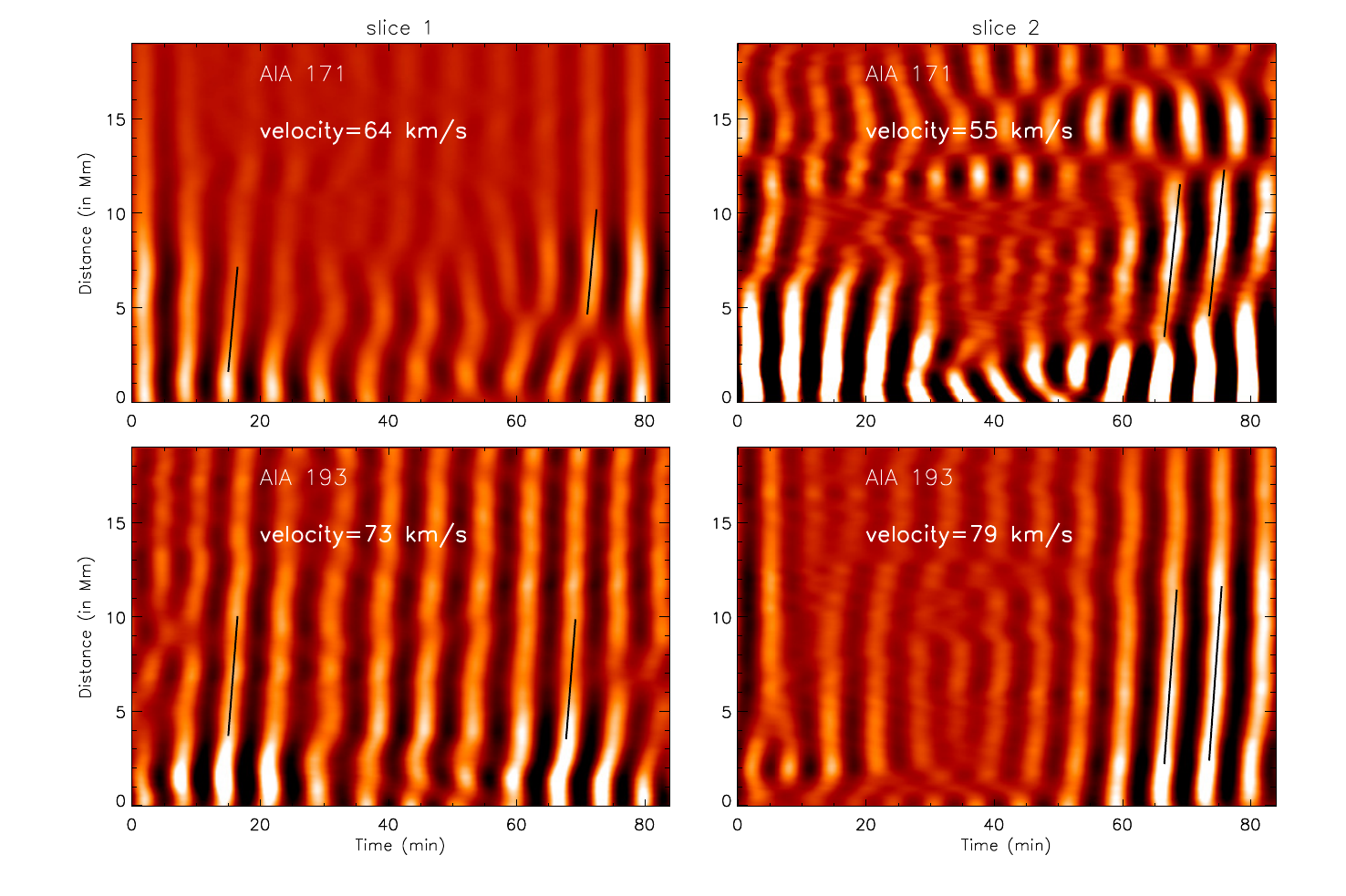}
\caption{\textrm{(Top Panel)}: {\it Left}: X-t map at slice 1 position and {\it Right}: X-t map at slice 2 position as shown in Fig.~\ref{fig1} (c) using 7 min Fourier filtered AIA 171~\r{A} image overplotted with best fit straight line using the method mentioned in section~\ref{imagesection}. \textrm{(Bottom Panel)}: Same as top panel for AIA 193~\r{A}. }
\label{fft} 
\end{figure*}

Since the periodicity $\sim$7 min is present in slices 0, 1 and 2, we create time-distance maps using Fourier filtered images. We perform Fourier transform at each pixel location. Fourier power is, then, multiplied with a Gaussian peaking at 7 min with FWHM of ~1 min and inverse Fourier transform is applied to get the reconstructed light curve at the same pixel location. Fourier filtered movies of AIA 171~\r{A} and 193~\r{A} are available online as movie 3 and 4 respectively. Time-distance maps for slices 1 and 2 using Fourier filtered images are shown in Fig.~\ref{fft}. We find that the periodicity of $\sim$7 min is significant in AIA 193~\r{A} and mostly towards the end of observation (from 60 min to 80 min in Fig.~\ref{fft} lower right panel) at slice 2 position. This is also evident from wavelet plots as shown in Fig.~\ref{wavelet} where we note that at slice 2 position 7 min periodicity is present at the start and towards the end of the observation from 60 to 80 min (see Fig.~\ref{wavelet} (d)). This fact is also supported by Fourier filtered movie 4 where we see the significant PDs propagating outwards at slice 2 position towards the end of observation. A Similar behaviour is found in AIA 171~\r{A} wavelet plots in slices 1 and 2 but the presence of the 7 min periodicity at later times is most prominent in AIA 193~\r{A} at slice 2 position. This indicates that these PDs are triggered by some drivers at specific instances.\\
Since significant PDs are observed in later times in slice 2 positions, we fit the significant ridges with a straight line between 60 and 80 min in AIA 171~\r{A} and 193~\r{A} Fourier filtered time-distance maps for slit 2 as shown in Fig.~\ref{fft} right panel. The method of fitting is the same as explained in Section~\ref{imagesection}. We estimate the velocity to be 64$\pm$3 km $s^{-1}$ (73$\pm$3 km $s^{-1}$) and 55$\pm$3 km $s^{-1}$ (79$\pm$2 km $s^{-1}$) in AIA 171~\r{A} (193~\r{A}) for slice 1 and 2 respectively. We find that 7 min period PDs are propagating with larger velocity in hotter channel (AIA 193~\r{A}) which further supports the fact that these could be propagating slow magnetoacoustic waves. However we would like to point out that the 7 min periodic PDs are not so clearly observed in AIA 171~\r{A} and we see uneven ridges which could affect the velocity estimation. In slice 1 position (Fig.~\ref{fft} left panel) 7 min periodic PDs are not clearly seen maybe because the 7 min period is not the dominant one in the time series as shown in Fig.~\ref{wavelet} (a). Thus we could fit only two ridges, one at 15 min and another at 65 min after the start of observation. We fit the corresponding ridges in AIA 193~\r{A} and find that in slice 1, too, the velocity is higher in hotter channel.



\subsubsection{Jet-like features in SJI 1300~\r{A}}
In this subsection we study the dynamic properties of the observed jet-like features. To determine the apparent speed of jets seen in network boundary we place nine artificial box slices five pixel wide in SJI 1330~\r{A} as shown in Fig.~\ref{fig1} (a). Slices are placed based on the direction of propagation of significant jets as seen in SJI 1330~\r{A} in Movie~2 (available online). We average along the width of each box slice to increase signal to noise. Corresponding time-distance maps are shown in Fig.~\ref{fig7}. In time-distance maps we fit significant ridges with straight lines. We find that several jets are preceded by brightenings seen in time-distance maps which points towards the fact that small scale reconnections could be triggering these jet-like features. We identify 62 jet-like features and fit them with dashed green curves as shown in Fig.~\ref{fig7}. The slope of the overplotted green dashed line gives an estimate of the apparent speed of the jet projected in the plane of sky. The velocity distribution of these jet-like features is shown in Fig.~\ref{fig7} (last panel). We find that the distribution peaks around 10 km~s$^{-1}$, which is similar to what has been reported by \citet{2007Sci...318.1591S}. Apart from outflows we see several downflows in the time distance maps, which suggest that a certain amount of the jet material is falling back (see slit 4 in Fig.~\ref{fig7}).

\subsubsection{Correspondence between Jet-like features in SJI 1300~\r{A} and PDs in AIA 171~\r{A} and 193~\r{A}}
\begin{figure*}
\centering
\includegraphics[angle=0,scale=.8]{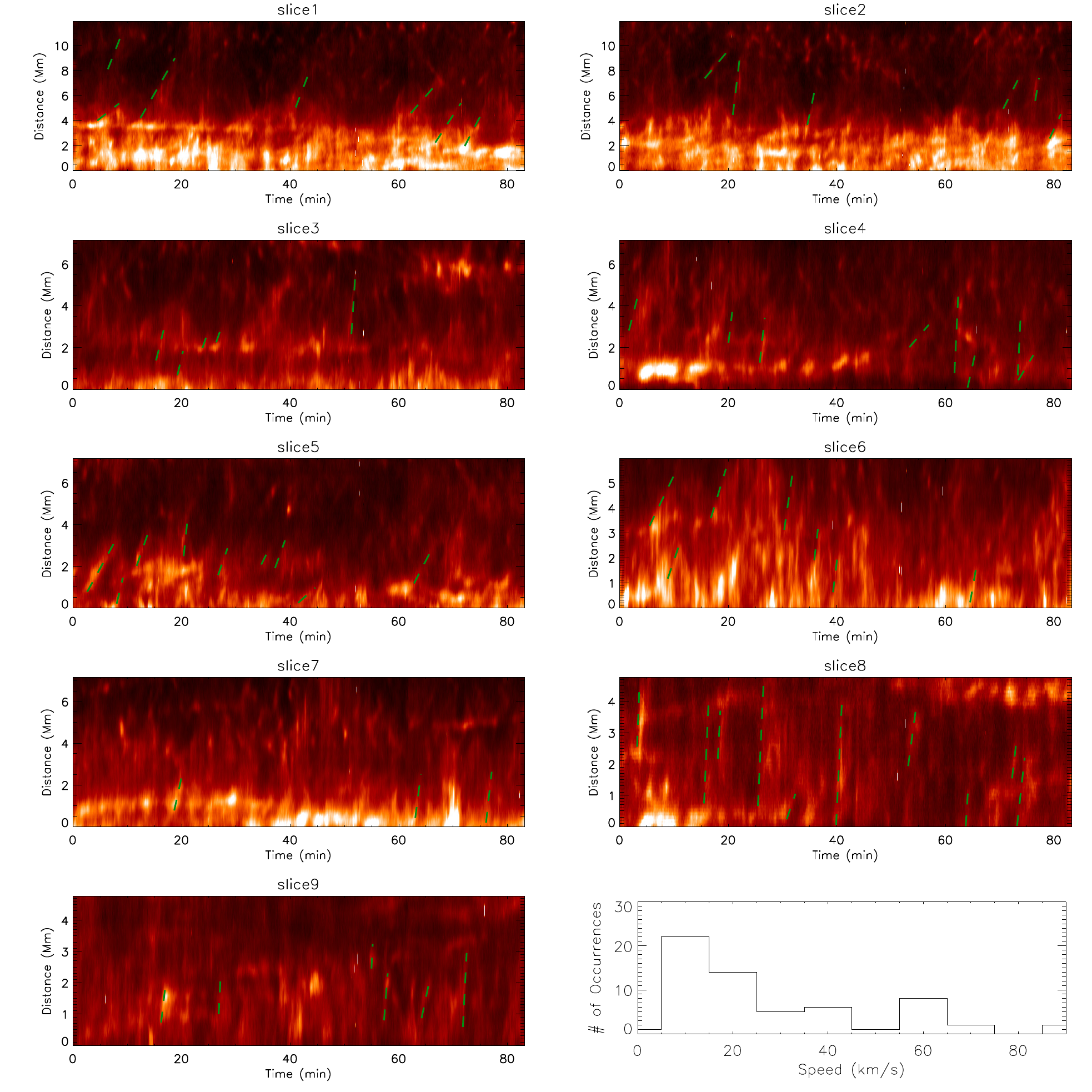}
\caption{X-t maps for 9 boxes slices as shown in Fig.~\ref{fig1} (a). The significant ridges in x-t maps are overplotted with dashed straight lines marked in green. The slope of the ridges gives an estimate of the velocity of the outward moving features (jet-like features). The velocity distribution of jet apparent speeds is shown in the last panel. The distribution peaks at 10 km$s^{-1}$.}
\label{fig7} 
\end{figure*}

\begin{figure*}
\includegraphics[scale=.6]{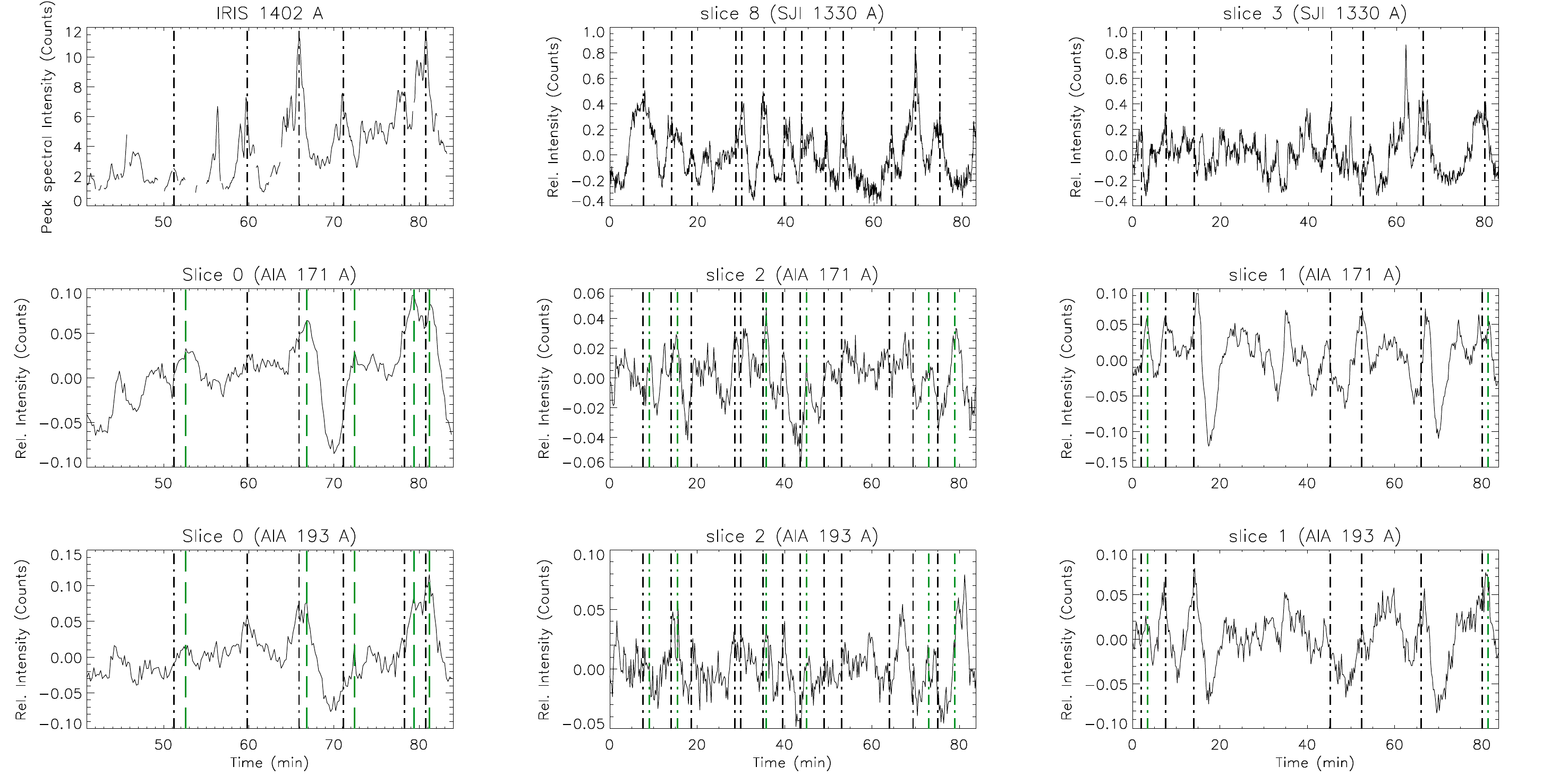}
\caption{\textrm{(Left Panel)}: {\it  Top}: IRIS spectral intensity at the position marked as trangle in Fig.~\ref{fig1}. {\it Middle}: Light curve of AIA 171~\r{A} at the same position. {\it Bottom}: Same as middle for AIA 193~\r{A}. Black dot-dashed lines correspond to peak in spectral intensity. Green dot-dashed lines mark the corresponding peaks in AIA 171~\r{A} and 193~\r{A}. The time lag can be estimated by estimating the time difference between the black and green lines. \textrm{(Middle Panel)}: {\it Top}: IRIS SJI 1330~\r{A} light curve at 0.5 Mm for x-t map created using box slice 8 as shown in Fig~\ref{fig7} (a). {\it Middle}: AIA 171~\r{A} light curve at 3 Mm (co-spatial with the position of light curve of Top Middle panel ) for x-t map created using box slice 2 in AIA 171~\r{A} as shown in Fig.~\ref{fig5} (top right panel). {\it Bottom}: AIA 193~\r{A} light curve at 3 Mm for x-t map created using slice 2 as shown in Fig.~\ref{fig5} (bottom right panel). \textrm{(Right Panel)}: Same as Middle Panel at position co-spatial with slice 3 as shown in IRIS SJI and slice 1 in AIA 171~\r{A} and 193~\r{A}.}
\label{sjipd} 
\end{figure*}

In this subsection we explore if the jet-like features can be responsible for the generation of the PDs.\\
To understand the source of PDs generated in the corona, we compare the spectral intensity of IRIS slit spectra at the foot point (marked as triangle in Fig.~\ref{fig1}) and the intensity of PDs observed in time-distance map of slit 0 (see Fig.~\ref{fig5}) at 2 Mm which coincides with the position of the triangle marked in Fig.~\ref{fig1}, for AIA 171~\r{A} and AIA 193~\r{A}. The left panel of Fig.~\ref{sjipd} compares the peak intensity of IRIS slit spectra (Top left), PDs (along slit 0) observed in AIA 171~\r{A} (middle left) and PDs (along slit 0) observed in AIA 193~\r{A} (bottom left). The dot-dashed line in black represents the peak spectral intensity while the green line represents the corresponding intensity peaks in AIA 171~\r{A} and 193~\r{A}. We find that IRIS spectral intensity peaks precede the AIA intensity peaks and we estimate a lag of 24 s -- 84s with a mean lag of $\sim$ 60 s.\\
Some of the peaks in spectral intensity do not correspond to sharp peaks in AIA 171~\r{A} and 193~\r{A}, maybe because the jet is aligned sideways and do not move along the slice 0 placed co-spatially with IRIS slit. We also compare the light curve at 0.5 Mm at box slice 8 placed in SJI 1330~\r{A} (see fig.~\ref{fig1}a) and 3 Mm at box slice 2 placed in AIA 171~\r{A} and 193~\r{A} (see Figs.~\ref{fig1} (c) and (d)). These two positions are also co-spatial.\\ 
From Fig.~\ref{sjipd}, we note that there is fairly good one-to-one correspondence between jet-like features in slice 8 in IRIS SJI and PDs in slice 2 in AIA 171~\r{A} and 193~\r{A}. The dot-dashed line represents the peak in IRIS SJI intensity. We also note that there is no significant lag except at the last ridge (marked in dot-dashed green line) where the lag is $~\sim$ 120 s. We carry out a similar analysis for co-spatial positions at slice 3 placed in SJI 1330~\r{A} and at slice 1 placed in AIA 171~\r{A} and 193~\r{A}. We found a lag of 84 s at two positions marked with green dot-dashed lines. \\
It should be noted that the time lag between peaks in IRIS SJI and AIA is not uniform. Apart from the fact that some jets are preceded by the brightenings in x-t maps, at some positions we may not capture the exact time when a jet starts in IRIS SJI (see, slice 8 in Fig.~\ref{fig7} between 40 and 80 min). What we know is when a jet appears in the position of slices placed in IRIS SJI. It is also possible that when jets appear in a given slice, they have already covered some distance in vertical direction (projection effect) and have generated the PDs visible in coronal channels. Thus they may appear co-temporal with PDs observed in AIA 171~\r{A} and 193~\r{A}.\\
These results allow us to propose that jet-like features in transition region may cause PDs as observed in the corona.\\





\section{Summary and Conclusions}
We study the dynamics at the foot points of an on-disk plume adjacent to a coronal hole using combined imaging, spectroscopic and magnetic measurements. We find that there is continuous emergence of new magnetic flux of positive polarity (see Movie~1 online) and there is fairly good correspondence between intensity enhancement and the  positive magnetic flux cancellation. 
This suggests that the emerging flux interacts with the existing fields, which results in reconnection and cancellation of the flux resulting in small jet-like outflows termed as jetlets as observed in AIA by \citet{2014ApJ...787..118R}. They conjectured that such small-scale jets are the consequences of flux cancellation at the foot points of plume. We think that jet-like features observed by \citet{2014ApJ...787..118R} are the PDs that we observe in AIA channels. They associated the jetlets seen in AIA channels to the underneath magnetic flux change. In this article we report that underneath magnetic flux changes generate jet-like features, which triggers PDs in the corona. These PDs can be explained in terms of magneto-acoustic waves. The small-scale jets are quite evident in the IRIS 1330 \r{A} SJIs (Movie~2 online) and the signature of reconnections are also prominent in the evolution of the line profiles. We observe enhanced line profile asymmetry, enhanced line width and large deviation from the average Doppler shift at specific instances. Association of plumes with supergranular network boundaries has been studied by several authors. In this study we find that plume foot points are coinciding with network boundaries as seen in IRIS 1330 \r{A} SJI. The observations of the origin of these jet-like features in connection with plumes is still lacking. In this article, we report the observational evidence of the origin of jet-like features from network boundaries thus confirming the earlier predictions. Footpoints of plumes are just enhanced network boundaries where we often see clear network jets in IRIS SJIs \citep{2014Sci...346A.315T}. IRIS spectra at plume foot point shows the presence of more than one Gaussian components, enhanced wings, high Doppler shifts and large fluctuations in intensity. This confirms the presence of flows \citep{2011ApJ...727L..37T, 2012ApJ...759..144T} at the foot point of the plume. These jet-like features load mass to plumes.\\
We estimate the speed of the jet-like features by placing several slices in IRIS SJI. These jet-like features originate from network boundaries due to small scale magnetic reconnections. We find that the speed distribution peaks at 10 km$s^{-1}$ which is the typical velocity for a chromospheric anemone jets as reported by \citet{2007Sci...318.1591S}. Thus most of the jets could be chromospheric in origin. Recently, \citet{2015ApJ...799L...3R} have reported jets of similar velocity range seen in IRIS SJI 1330~\r{A} to be associated with rapid blue- or red-shifted excursions seen in H-$\alpha$ images.  We find speeds of jets that are far lower than 150 km~$s^{-1}$ reported in \citet{2014Sci...346A.315T}. The datasets used by \citet{2014Sci...346A.315T} were centered on coronal holes while in this study the plume is located in quiet Sun region. The open field lines in coronal holes may accelerate the jets to larger velocities. Apart from this, the effect of projection of jets in plane of sky can not be ruled out.\\
We compared IRIS spectral intensity with PDs observed in AIA 171~\r{A} and 193~\r{A} and find that at several time instances the intensity enhancement in IRIS spectra precedes PDs observed in AIA channels. We found the lag of 24 s to 84 s with a mean lag of 60~ s. We also compared jet-like features in IRIS SJI with PDs observed in AIA channels at two different slice positions and find fairly good correspondence. However, at some instances we could not find one-to-one correspondence between IRIS SJI jet-like features and PDs observed in AIA channels. One reason could be that if jets are short lived then they appear in IRIS SJIs but PDs in corona may not be observed. We should point out that IRIS SJI 1330~\r{A} has strong contribution from C II which forms at upper chromosphere. Thus IRIS SJI 1330~\r{A} looks at upper chromosphere and lower transition region. While emission in AIA 171~\r{A} and 193~\r{A} represents the million degree plasma. Thus if reconnection happens at larger heights (upper transition region or lower corona) and the plasma is heated to coronal temperatures then we may see faint jets in IRIS SJI but strong PDs in AIA channels. Under this scenario our observed jets may not propagate to higher heights and they may not be responsible for the generation of  PDs in the corona. In the future, one can address this issue with better coordinated observations at several heights from chromosphere to corona. Therefore, we propose that there is some observational  evidence that jet-like features in network boundaries may cause the generation of PDs in corona.\\
Time-distance maps for AIA 171 \r{A} and 193 \r{A} reveal that the speed of quasi-periodic propagating intensity disturbances is larger in the hotter channel (AIA 193 \r{A}) than in AIA 171 \r{A}. Therefore, we believe that these are propagating slow magnetoacoustic waves. \citet{2012ApJ...754..111O} reported that the impulsive onset of flows with subsonic speed results in excitation of damped slow magnetoacoustic waves in active region loops. \citet{2013ApJ...775L..23W} performed 3D MHD modelling of active region fan loops and report that the PDs observed in coronal fan loops are due to several tiny upflows caused by nanoflares at the loop footpoints.  \citet{2011ApJ...737L..43N} reported the presence of flows close to the foot point and upward propagating slow magnetoacoustic waves at higher locations using EIS on board {\it Hinode}. Thus it appears that any form of reconnection outflow may generate slow waves in an expanding loop system. In this study, we report that the periodicities of intensity disturbances found in the corona using AIA 171 \r{A} and 193 \r{A} match well with periodicities found in IRIS spectral intensity and in IRIS SJI intensity in the transition region. Therefore, some jets observed in the transition region may be connected to PDs observed in the corona. This fact, together with the good correspondence between some jet-like features in IRIS images and some PDs in AIA coronal passbands, allows us to propose that the small jet-like upflows (jetlets) at lower heights (Transition region) are the drivers of slow magnetoacoustic waves seen at coronal heights in plumes. 

\acknowledgments
We thank the IRIS team for providing the data in the public domain. We thank Dr.~Hui Tian for his valuable suggestions which has enabled us to improve the manuscript. We also thank the anonymous referee for  careful reading and constructive suggestions. IRIS is a NASA small explorer mission developed and operated by LMSAL with mission operations executed at NASA Ames Research center and major contributions to downlink communications funded by the Norwegian Space Center (NSC, Norway) through an ESA PRODEX contract. We acknowledge a partial support from the Belgian Federal Science Policy Office through the ESA-PRODEX program.

\bibliographystyle{apj}
\bibliography{references}


\end{document}